\documentclass[useAMS,usenatbib]{mn2e}
\usepackage{amsmath}
\usepackage[pdftex]{graphicx}
\usepackage{epstopdf}
\usepackage[english]{babel}
\usepackage{subfigure}
\usepackage{array,multirow}
\usepackage{underscore}
\usepackage{natbib}

\usepackage{array}
\newcolumntype{L}[1]{>{\raggedright\let\newline\\\arraybackslash\hspace{0pt}}m{#1}}
\newcolumntype{C}[1]{>{\centering\let\newline\\\arraybackslash\hspace{0pt}}m{#1}}
\newcolumntype{R}[1]{>{\raggedleft\let\newline\\\arraybackslash\hspace{0pt}}m{#1}}

\def\mnras{MNRAS}
\def\apj{ApJ}
\def\aj{AJ}

\def\aaps{A\&AS}
\def\apjl{ApJL}
\def\apjs{ApJS}

\def\nat{Nature}

\title[M87 from parsec to megaparsec scales]{Galaxy structure from multiple tracers: II. M87 from parsec to megaparsec scales}
\author[L. J. Oldham and M. W. Auger]{L. J. Oldham\thanks{E-mail: loldham@ast.cam.ac.uk}, M. W. Auger
\\
Institute of Astronomy, University of Cambridge, Madingley Road, Cambridge CB3 0HA, UK}

\pubyear{2015}

\begin{document}
\maketitle
\setcounter{page}{1}

\begin{abstract}

Following a number of conflicting studies of M87's mass profile, we undertake a dynamical analysis of multiple tracer populations to constrain its mass over a large radius range. We combine stellar kinematics in the central regions with the dynamics of 612 globular clusters out to 200~kpc and satellite galaxies extending to scales comparable with the virial radius. Using a spherical Jeans analysis, we are able to disentangle the mass contributions from the dark and baryonic components and set constraints on the structure of each. Assuming isotropy, we explore four different models for the dark matter halo and find that a centrally-cored dark matter distribution is preferred. We infer a stellar mass-to-light ratio $\Upsilon_{\star,v} = 6.9 \pm 0.1$ -- consistent with a Salpeter-like IMF -- and a core radius $r_c = 67 \pm 20$~kpc. We then introduce anisotropy and find that, while the halo remains clearly cored, the radial stellar anisotropy has a strong impact on both $\Upsilon_{\star,v}$ and the core's radius; here we find $\Upsilon_{\star,v} = 3.50_{-0.36}^{+0.32}$ -- consistent with a Chabrier-like IMF -- and $r_c = 19.00_{-8.34}^{+8.38}$~kpc. Thus the presence of a core at the centre of the dark halo is robust against anisotropy assumptions, while the stellar mass and core size are not. We are able to reconcile previously discrepant studies by showing that modelling the globular cluster data alone leads to the very different inference of a super-NFW cusp, thus highlighting the value of multiple-population modelling, and we point to the possible role of M87's AGN and the cluster environment in forming the central dark matter core.

\end{abstract}


\section{Introduction}

The $\Lambda$CDM paradigm of structure formation has been very successful in describing the Universe on large scales, but there remains some tension regarding its predictions about galaxy structure. For instance, one main prediction of cold, collisionless gravitational collapse is the formation of a central cusp in the density profile of the dark matter (DM) haloes that envelope galaxies, with $\rho_{DM} \sim r^{-\gamma}$ and $\gamma = 1$  at small radii \citep{NFW1997, Navarro2010}. However, real haloes also contain baryons, and the imprint of baryonic physics on the DM distribution could be significant. For instance, feedback from supernovae and AGN, as well as dynamical friction from infalling satellites, could lead to some degree of heating and expansion \citep[e.g.][]{Mashchenko2006, Laporte2012,Governato2012, Velliscig2014}, thus hollowing out the DM, while the cooling and condensation of baryons could increase the density in the central regions via adiabatic contraction \citep[e.g.][]{Blumenthal1986, Gnedin2004}. The current observational picture reflects this complexity, with the haloes of an increasing number of systems being found to favour cored or only weakly cuspy central profiles. For instance, the recent local surveys THINGS and LITTLE THINGS \citep{Hunter2007} found a large fraction of dwarf field galaxies to have DM density profiles that go as $\rho \propto r^{-0.4}$ within the central kiloparsec, and studies of low-surface-brightness galaxies also point to relatively flat central profiles with a large scatter \citep[e.g.][]{deNaray2011}. 

While a great deal of progress has been made in constraining the halo structure of low-surface-brightness galaxies and dwarf spheroids, where DM dominates over the baryonic mass, the situation is much more complicated for their massive elliptical counterparts. Here, our ignorance about the stellar initial mass function (IMF) introduces a degeneracy between dark and luminous matter, which makes it hard to constrain the behaviour of the DM in the inner regions; equally, the task of probing the gravitational potential at large radii is made hard by the fact that their outskirts are notoriously faint. However, one way of significantly alleviating these degeneracies is to use multiple dynamical tracer populations, spanning a range of galactocentric radii \citep[e.g.][]{Schuberth2010, Walker2011, Napolitano2014}. Indeed, massive elliptical galaxies are often home to large populations of planetary nebulae (PNe), globular clusters (GCs) and even, in the case of brightest cluster galaxies (BCGs), satellite galaxies, and each of these populations, with its own signature spatial distribution and kinematic profile, can be used as an independent probe of the gravitational potential. The pool of ETGs for which such an analysis has been carried out is currently too small for any meaningful conclusions to be extracted, though the suggestion also appears to be that the NFW profile may not provide a good universal fit: studies of BCGs by \cite{Sand2004}, \cite{Newman2011} and \cite{Newman2013} have found evidence for sub-NFW density profiles in clusters, while a handful of studies of field ellipticals \citep[e.g.][]{Grillo2012, Sonnenfeld2015}, have found super-NFW densities. Clearly, a lot remains to be understood here, from both observational and theoretical perspectives.

The massive elliptical M87, located at the centre of the Virgo cluster, is an ideal subject for continuing such studies, as it has an enormous GC population \citep[estimated as $\sim$ 12,000, e.g.][]{McLaughlin1994, Tamura2006b, Oldham2015}, making it one of the richest GC hosts in the local Universe. Further, the sample of this population for which we have high-resolution kinematic data has been greatly expanded in recent years through the wide-field study of \cite{Strader2011}. It is also understood to be a slow rotator and nearly spherical \citep[e.g.][]{Cappellari2006}, indicating that it is suitable for mass modelling under the assumption of spherical symmetry \citep[though see also][for recent evidence for non-axisymmetry]{Emsellem2014}.  However, two recent studies of M87's mass distribution, both primarily based on Strader's GC dataset, are markedly inconsistent. The first of these, \cite{Agnello2014}, divided the GC sample into three independent populations and used a virial analysis to infer a very cuspy central density profile ($\gamma = 1.57$), while \cite{Zhu2014} combined SAURON central stellar kinematics with the Strader GC data (along with an additional GC sample from Hectospec) and modelled the density profile using a logarithmic potential, thus imposing a core, which they infer to be $r_s = 42 \pm 10$ kpc. Their total inferred stellar masses also differ by almost a factor of two. As M87 is one of the nearest and most well-observed BCGs, it seems unsatisfactory that its mass distribution should still be so poorly constrained, and clearly there remains work to be done. The aim of this paper is therefore to use a synthesis of GC, satellite and stellar kinematic data in conjunction with flexible mass models in order to infer a density profile which is free to be cuspy or cored as the dynamics dictate. 

The paper is organised as follows: in Section~\ref{sec:data}, we introduce the tracer populations used in our analysis and the associated datasets. In Section~\ref{sec:massmodel} we describe our mass model and Jeans analysis, the results of which are presented in Section~\ref{sec:results}. We discuss the implications of our findings in Section~\ref{sec:discussion}, and use Section~\ref{sec:conclusions} to summarise our main conclusions. Throughout this work, we assume a distance to M87 $D_L = 16.5$ Mpc.

\section{DATA}
\label{sec:data}

To constrain M87's density profile across a wide radius range, we use multiple dynamical tracers, combining stellar kinematics in the central regions with GC dynamics at large radii and satellite galaxies on cluster scales. We can then solve the Jeans equation for each population separately, provided that the underlying density distribution is known. We therefore take data from a number of sources, as summarised in Table 1.

\begin{table*}
\centering
\begin{tabular}{|c|cccc|}\hline
  & radial coverage & data type & instruments &  sources \\\cline{1-5}
\multirow{1}{*}{stars}   & $1.6 \times 10^{-3}$ - 210 kpc & photometry & multiple & \protect\cite{Kormendy2009} \\
  &  0 - 2.5 kpc & kinematics & \parbox{6cm}{\centering SAURON/William Herschel Telescope} & \protect\cite{Emsellem2004} \\
  & 0 - 0.17 kpc & kinematics & NIFS/Gemini Telescope & \protect\cite{Gebhardt2011} \\\hline
\multirow{1}{*}{GCs} & 1.3 - 240 kpc & photometry & MegaPrime/CFHT & \protect\cite{Oldham2015} \\
  & 2 - 200 kpc & kinematics & multiple & \protect\cite{Strader2011} \\\hline
\multirow{2}{*}{\parbox{1.5cm}{\centering satellite galaxies}} & 35 - 1000 kpc & kinematics & multiple (mostly SDSS) & \protect\cite{Kim2014} \\
 & & & & \protect\cite{Blakeslee2009} \\\hline
\end{tabular}
\caption{Various sources of photometric and spectroscopic data for the different tracer populations used in the dynamical analysis.}
\label{tab:sources}
\end{table*}

\begin{figure*}
  \centering
  \subfigure{\includegraphics[trim=20 0 10 0,clip,width=0.5\textwidth]{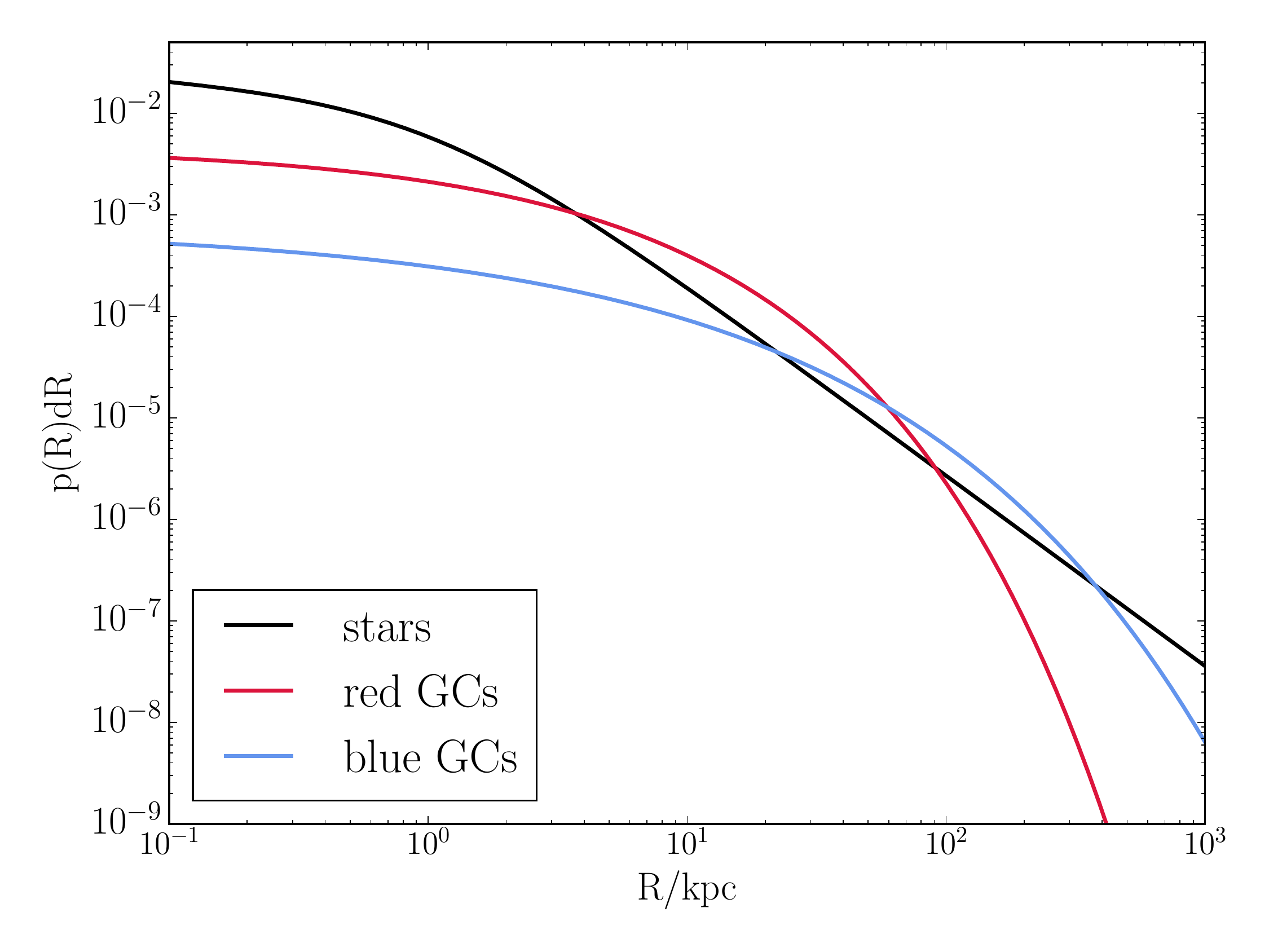}}\hfill
  \subfigure{\includegraphics[trim=20 0 10 0,clip,width=0.5\textwidth]{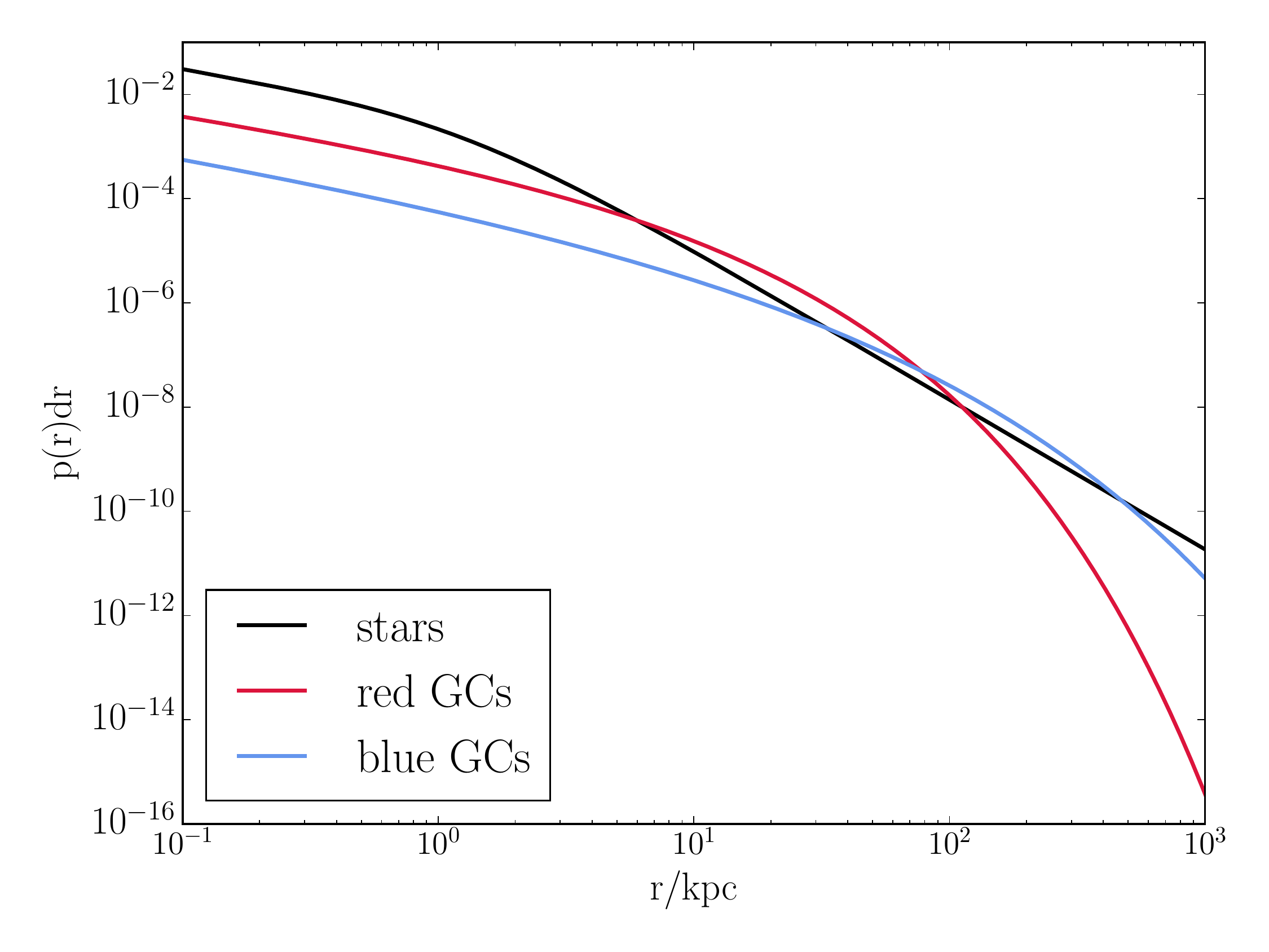}}\hfill
 \caption{Left: Normalised surface brightness profiles for the starlight and the red and blue GCs, scaled arbitrarily. The stellar surface brightness was obtained from a fit to the V-band profile presented in \protect\cite{Kormendy2009}, while the GC distributions come from the modelling in \protect\cite{Oldham2015}. Right: Normalised 3D luminosity density profiles for the same three populations, deprojected assuming spherical symmetry.}
 \label{fig:SBs}
\end{figure*}

\subsection{Stars}
\label{sec:data:stars}

The deprojected stellar surface brightness profile comes into our analysis at two points: first, the use of the stars as dynamical tracers in the Jeans equation requires us to know their 3D density distribution; second, our goal is to model M87's mass as the sum of dark and luminous components, and the latter is simply the product of the integrated 3D luminosity density with some constant mass-to-light ratio, $\Upsilon_{\star}$, which can be inferred from the data. We use the radial profile for M87 presented in \cite{Kormendy2009}, in which 20 sets of observations across a range of radii and filter systems were synthesised into a single profile in the V band. As M87 is known to have a very extended cD envelope in addition to a stellar core \citep[eg.][]{Chakrabarty2007}, its surface brightness profile cannot be accurately modelled by a S\'ersic profile at both small and large radii. We therefore chose to model it using a more flexible Nuker profile, according to the following relation

\begin{equation}
 I(R) = I_0 \Big(\frac{r}{r_b}\Big)^{-\zeta} \Big(1 + \Big[\frac{r}{r_b}\Big]^{\alpha}\Big)^{\frac{\zeta-\eta}{\alpha}}
\end{equation}
 
with amplitude $I_0$, break radius $r_b = 1.05$ kpc, inner slope $\zeta = 0.186$, outer slope $\eta = 1.88$ and break softening $\alpha = 1.27$. While this has the disadvantage of having no analytic deprojection or normalisation, it is much more flexible than a S\'ersic profile as it allows both the inner and outer slopes greater freedom. Assuming  spherical symmetry, we deproject this profile to give the 3D density shown in Figure~\ref{fig:SBs}.

The kinematics of the inner 33$'' \times $41$''$ of M87 have been observed with the integral-field unit (IFU) SAURON, and a catalogue of the first four moments of the Gauss-Hermite expansion of the line-of-sight velocity distribution is available online\footnote{http://www.strw.leidenuniv.nl/sauron/}. We use the velocity dispersions, which were obtained from the spectra using a direct pixel fitting routine \citep[e.g.][]{vanderMarel1994}. The spectra were adaptively binned to ensure a signal-to-noise of at least 60 per spectral resolution element; uncertainties are generally less than $\sim$ 20 kms$^{-1}$, with a mean uncertainty of $\sim 9 \text{kms}^{-1}$. 

M87 is known to host a supermassive black hole (SMBH) of mass $\sim 6.6 \times 10^9 M_{\odot}$ \citep{Gebhardt2009,Gebhardt2011}, and this should make a significant contribution to stellar velocity dispersions at the smallest radii. The SAURON dataset extends right down to the centre, though its resolution of 1$''$ is too low to be able to set constraints on the SMBH mass. As we are mainly interested in the behaviour of the DM and stellar components, one option to deal with this would be to simply exclude the apertures within the central $\sim 3 ''$ from our analysis. However, though the contribution of the SMBH to the enclosed mass becomes sub-dominant beyond this approximate radius, it still contributes non-negligibly at larger radii and may be covariant with the stellar mass. We therefore include the SMBH in our mass model and constrain it using high-resolution kinematics of the central 2$''$, as observed with the IFU NIFS on the Gemini North Telescope \citep{Gebhardt2011}. As explained in that paper, these data were obtained from spectra which used laser adaptive optics corrections, and have a resolution of 0.08$''$ and a signal-to-noise generally greater than 50. The velocity moments are provided in radius and position angle bins, though our simplifying assumption of spherical symmetry allows us to combine bins azimuthally.

\begin{figure*}
  \centering
  \subfigure{\includegraphics[trim=10 0 15 0,clip,width=0.5\textwidth]{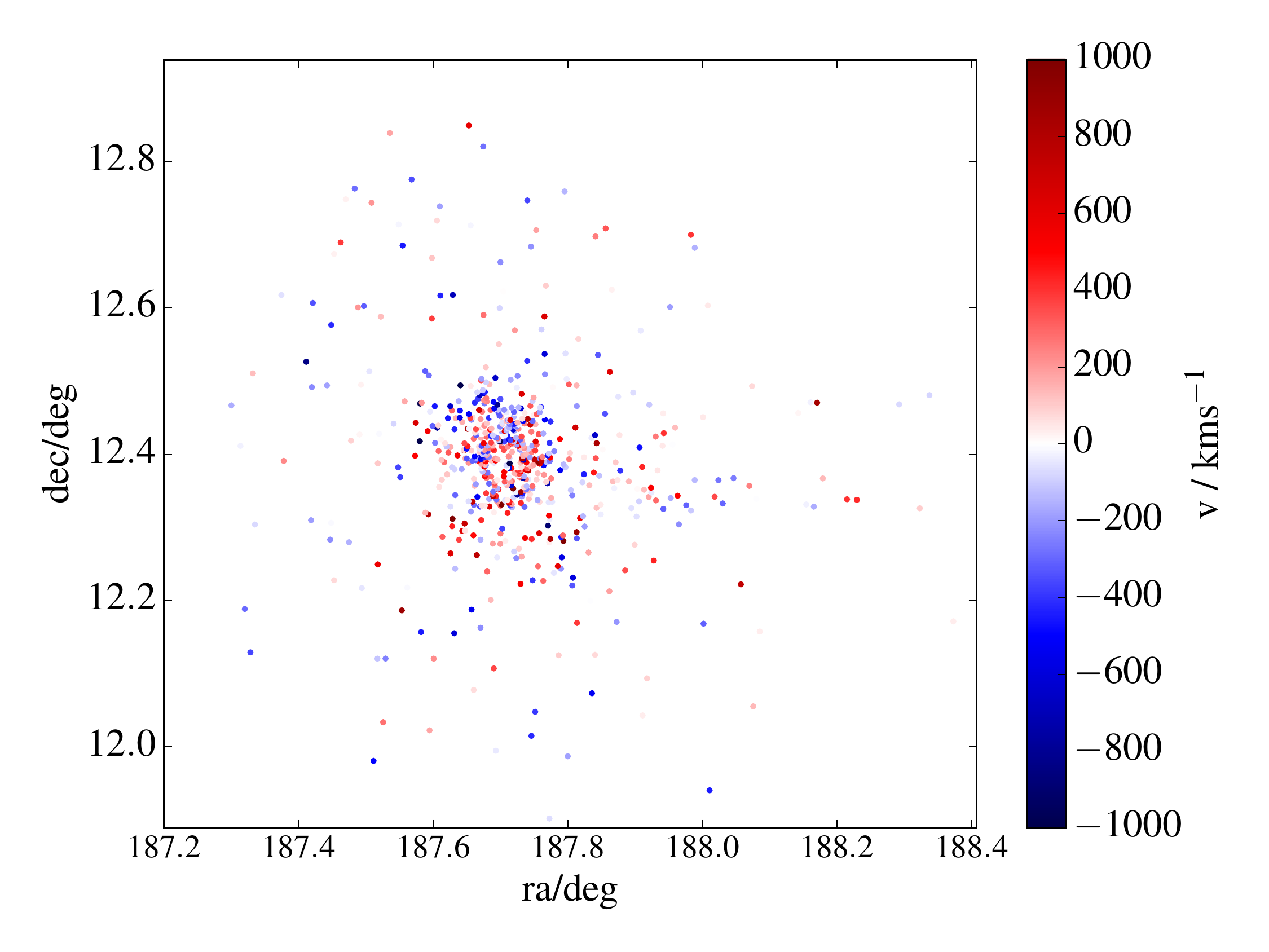}}\hfill
  \subfigure{\includegraphics[trim=15 0 15 10,clip,width=0.5\textwidth]{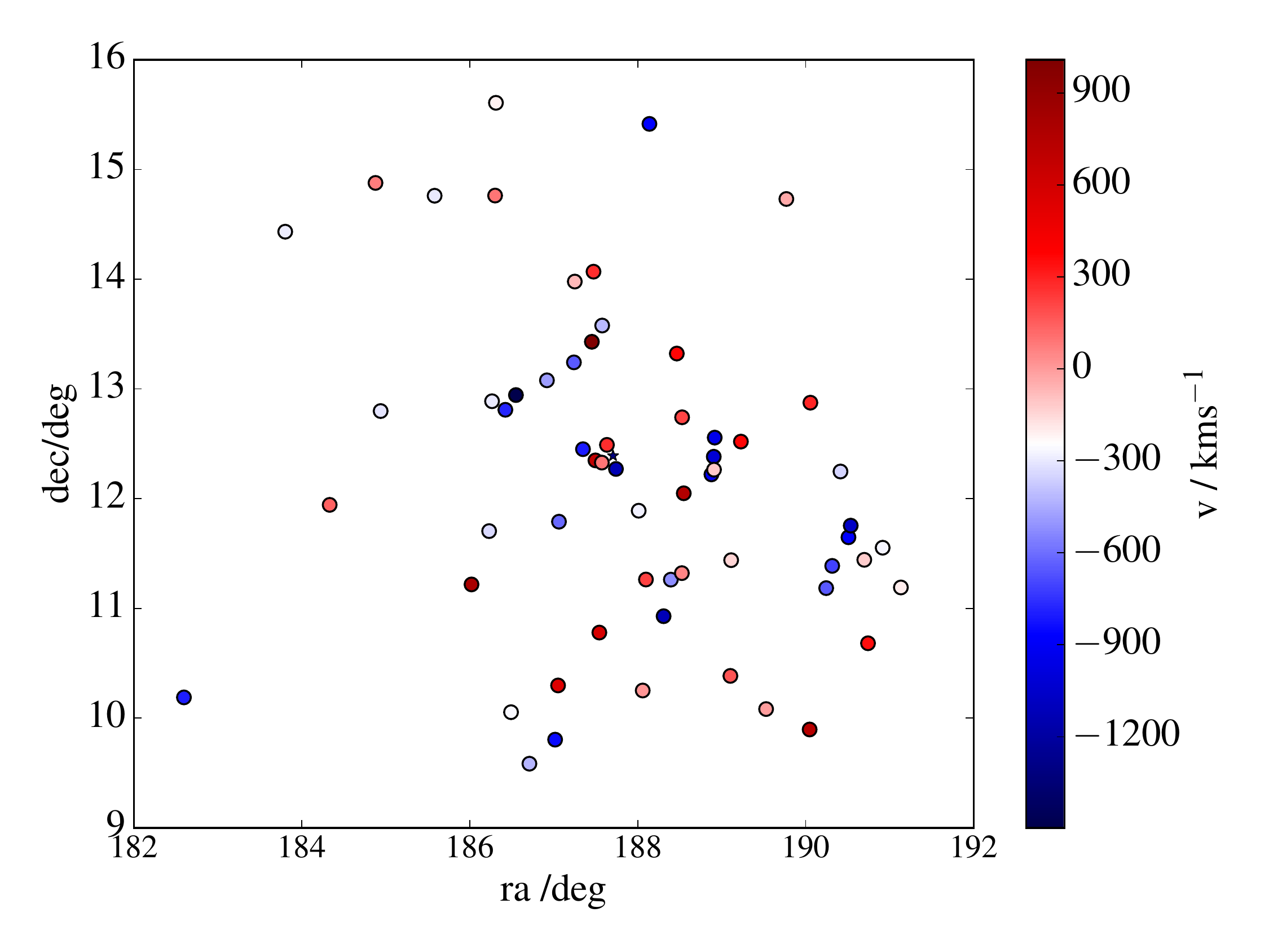}}\hfill
\caption{Maps of the tracer populations, coloured by velocity with respect to M87. Left: kinematic GC sample. Note that this does not follow the distribution of the underlying population, which is characterised independently using the photometric catalogue of Oldham \& Auger (2015, accepted). Right: satellite galaxy sample. M87 itself is plotted as a navy star. Note that we do not expect this spectroscopically-selected sample to be complete.}
\label{fig:GCmap}
\end{figure*}

\subsection{Globular clusters}
\label{sec:data:GCs}

We use the colour and radial profiles for the GC populations of \citet{Oldham2015}, as shown in Figure~\ref{fig:SBs}. The details of the inference are explained fully in that paper, but in brief, archival CFHT/MegaPrime images in the \textit{ugriz} bands were used to compile a sample of 17620 GC candidates, selected according to their colours, magnitudes and apparent sizes, and the resulting catalogue was modelled to infer the radial and colour distributions and the relative fractions of the two GC populations. The surface density for each GC component was modelled as a S\'ersic profile

\begin{equation}
 N(R) = N_0 \exp{{\Bigg[-k_n\Big(\frac{R}{R_e}\Big)^{\frac{1}{n}}\Bigg]}}
\end{equation}
with $k_n = 2n - 0.324$, and with the profiles normalised such that $\int_0^{R_{max}}N(R)dR = 1$, where $R_{max} = 240$ kpc is the radius of the outermost GC in the catalogue. The $g-r$, $g-i$ colours and the GC luminosity function for each GC population were modelled as Gaussians, with radius-dependent colour gradients in all but the latter. The 3D deprojected profiles are also shown in Figure~\ref{fig:SBs}.

\cite{Strader2011} presents a spectroscopic catalogue of 737 GC candidates around M87 at radii from 2 kpc to 200 kpc (plus one object at 800 kpc, which we exclude because it lies outside the region over which our model from the photometry is strictly valid). The catalogue combines new measurements for 451 GCs -- obtained using Keck/DEIMOS, Keck/LRIS and MMT/Hectospec -- with literature data, and provides a `classification' of objects as blue GCs, red GCs, ultra-compact dwarfs and transient/unknown along with SDSS-band $g-r$ and $g-i$ colours. We cross-correlate our photometric catalogue with this spectroscopic catalogue, selecting only the objects classified in \citet{Strader2011} as GCs, to obtain a sample of 612 GCs with complete luminosity, spatial, colour and kinematic information. In the analysis that follows, we choose to use the \cite{Oldham2015} photometry over that provided in \citet{Strader2011}, for consistency with our GC colour distributions.

\subsection{Satellite galaxies}
\label{sec:data:satellites}

The Extended Virgo Cluster Catalogue \citep[EVCC,][]{Kim2014} provides the redshifts and positions on the sky of 1589 galaxies in a footprint of 725 deg$^2$ centred on M87, extending to 3.5 times the cluster virial radius. The redshifts are compiled from the SDSS DR7 release and the NASA Extragalactic Database (NED), and each object is classified as either a certain cluster member, a possible  member or a background source based on morphological and spectroscopic criteria. As it is important that our sample only contains galaxies moving in M87's halo potential, we selected only those objects classed as certain members according to both criteria, and we further cross-correlated these with the catalogue of \cite{Blakeslee2009}, which used surface brightness fluctuations to calculate distance moduli. To avoid contamination from the W cloud, a slightly more distant component of the Virgo cluster at a characteristic distance of 23 Mpc, we imposed a distance cut of 20 Mpc; further, to separate the A cloud (centred on M87) from the B cloud (centred on M49) we also imposed a declination angle cut of 9.5 degrees and a radius cut of 1 Mpc. The spatial and velocity distributions of the resulting sample are shown in Figure~\ref{fig:GCmap}: it comprises 60 galaxies, with radii relative to M87 ranging from 35 kpc to 1 Mpc. 

\begin{figure}
 \centering
  \includegraphics[width=0.5\textwidth]{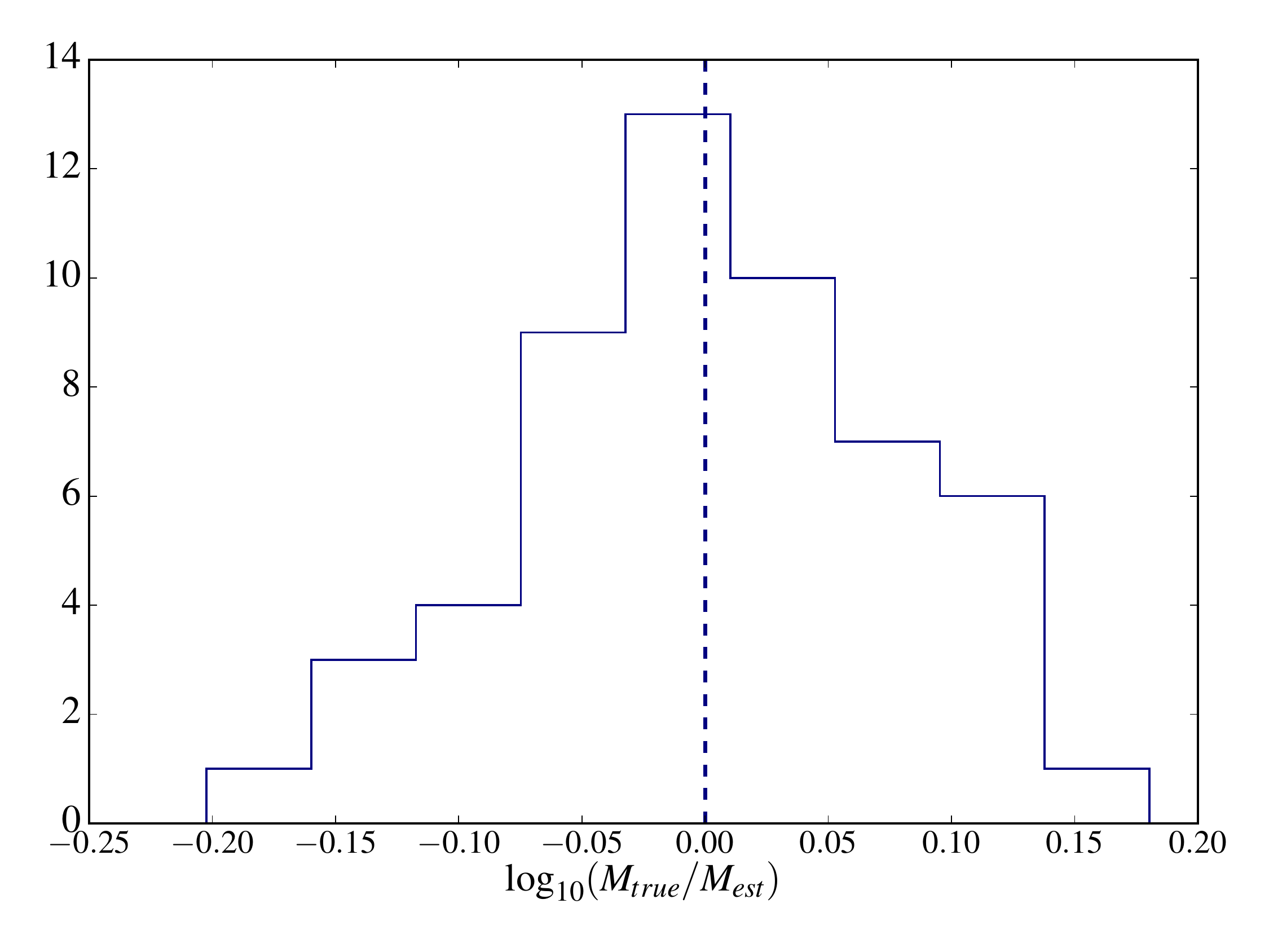}
  \caption{The ratio of the true halo mass to that estimated from the virial mass estimator, as calibrated from the simulations. The spread in estimates reflects the fact that the parameters $\beta$, $\mu$ and $\nu$ do in fact vary between the different groups, and are not truly global as we have assumed. However, the scatter is small at 0.1 dex and the median offset is negligible, indicating the robustness of the estimator.}
\label{fig:VME}
\end{figure}

We could use this tracer population in the same way as the stars and the GCs and require its velocity dispersion profile to satisfy the Jeans equation so as to obtain a further probe of the mass at scales comparable to the virial radius. However, as noted earlier, the Jeans equation requires us to know the tracer density distribution and, as our satellite sample is most likely highly incomplete, we do not have access to this quantity. Most importantly, the sample of satellites we use has been selected spectroscopically, and this imposes a non-trivial selection function which may alter the spatial distribution from the true underlying one, leading us to draw incorrect conclusions from a Jeans analysis. Instead, we choose to use these satellites to give an estimate of the total mass at large radii and so give a further constraint in our inference on the mass profile. We do this using the virial mass estimator developed in \cite{Watkins2010} and applied to a galaxy group in \cite{Deason2013}, which is designed to be robust against simple approximations to the true distributions. We use the mass estimator given by

\begin{equation}
M(R<R_{out}) = \frac{C}{G}\big<v_{los}^2 r^{\mu}\big>,
\end{equation}
where 

\begin{equation}
 C = \frac{\mu + \nu - 2\beta}{I_{\mu,\nu}}r_{out}^{1-\mu}
\end{equation}
and

\begin{equation}
 I_{\mu,\nu} = \frac{\pi^{0.5}\Gamma(\frac{\mu}{2} + 1)}{4\Gamma(\frac{\mu}{2} + \frac{5}{2})}[\mu + 3 - \beta(\mu + 2)].
\end{equation}
Here $\beta$ is the anisotropy parameter 

\begin{equation}
\beta = 1 - \frac{\sigma_t^2}{\sigma_r^2}
\label{eq:binney}
\end{equation}
for radial and tangential velocity dispersions $\sigma_r^2$ and $\sigma_t^2$, and $\mu$ and $\nu$ are the slopes of the potential and tracer density respectively, both assumed to be scale-free. As mentioned previously, our satellite sample is likely to be incomplete, and this means we cannot infer $\beta$, $\mu$ and $\nu$ directly from the data; instead we calibrate their values using simulations. Following \citet{Deason2013}, we use the $z = 0$ halo catalogue of the first MultiDark simulation, desribed in detail in \cite{Prada2012}. This uses the WMAP5 cosmology and contains about 8.6 billion particles per Gpc/h$^3$: the halo finder uses the bound density maximum technique described in \cite{Klypin1997}. We identify all haloes with more than 30 subhaloes and treat each subhalo as a distinct satellite galaxy. We then use the subhalo velocities and positions and the parent halo mass profiles to infer the posterior distributions on $\beta$, $\mu$ and $\nu$ that best describe the global properties of the population, and use these to generate a posterior on the total mass $M(R<R_{out})$ of our cluster, whose median and standard deviation we use to re-select haloes from the simulation and iterate the procedure until our the inference on $\mu$, $\beta$ and $\nu$ converges. We report median values $\beta = 0.30$, $\mu = 0.15$, $\nu = 1.9$, and find a cluster mass

\begin{equation}
 \log(M(R<985\text{kpc})/M_{\odot}) = 14.11 \pm 0.19.
\end{equation}
To test the calibration, we use the median values of $\beta$, $\mu$ and $\nu$ to apply the virial mass estimator to the simulated parent haloes, and the resulting comparison is shown in Figure~\ref{fig:VME}. Encouragingly, the masses are consistent, with a negligible median offset and a scatter of 0.1 dex, thus illustrating the robustness of the estimator. The applicability of this to M87 is then dependent on the assumption that the properties of the simulated galaxy satellite populations are representative of real galaxies.

\section{Modelling}

\subsection{Mass models}
\label{sec:massmodel}
We model the galaxy density profile as the sum of luminous and DM components plus a SMBH:

\begin{equation}
 \rho(r) = \rho_{\star}(r) + \rho_{DM}(r) + \rho_{BH}(r).
\end{equation}
The stellar density profile is obtained by deprojecting the stellar surface brightness and scaling by the stellar mass-to-light ratio $\Upsilon_{\star}$, which we assume to be constant with radius and include as a free parameter in the model. To allow flexibility in the DM density profile and to explore the impact of changing the halo model on the mass inference, we carry out our analysis using four different halo profiles. The first is a standard NFW profile,
\begin{equation}
 \rho_{DM}(r) = \frac{\rho_0}{4\pi} \Big(\frac{r}{r_s}\Big)^{-1} \Big(1 + \frac{r}{r_s}\Big)^{-2},
\end{equation}
with scale radius $r_s$ and density scale $\rho_0$, and cumulative mass

\begin{equation}
 M_{DM}(r) = \rho_0 r_s^3\Big(\ln\big(1+\frac{r}{r_s}\big) - \frac{r}{r+r_s}\Big).
\end{equation}
Following \citet{Zhu2014}, we also use a cored isothermal profile with a logarithmic potential (the LOG model)

\begin{equation}
  \rho_{DM}(r) = \frac{\rho_{0}r_s^2}{4\pi} \frac{3r_s^2 + r^2}{(r_s^2 + r^2)^2}.
\end{equation}
This model is inherently cored, but in contrast to the NFW has a $\rho \sim r^{-2}$ dependence at large radii: for a given core, then, this profile allows more dark matter to be placed at large distances from the galaxy centre. Its cumulative mass is given by 

\begin{equation}
 M_{DM}(r) = \frac{\rho_0 r_s^2 r^3}{r_s^2+r^2}.
\end{equation}
We also use a generalised NFW (gNFW),

\begin{equation}
 \rho_{DM}(r) = \frac{\rho_{0}}{4\pi}\Big(\frac{r}{r_s}\Big)^{-\gamma} \Big(1 + \frac{r}{r_s}\Big)^{\gamma-3},
\end{equation}
with free parameters $\gamma, \rho_0$ and $r_s$, where $\gamma$ is the inner slope. This has the advantage of leaving the profile free to choose between cusps ($\gamma \geq 1$) and cores ($\gamma = 0$) in the centre, while still becoming NFW-like at large radii in a way that is consistent with both simulations and observations. This gives a cumulative mass profile that can be related to the Gauss hypergeometric function $_2F_1$ as

\begin{equation}
 M_{DM}(r) = \frac{\rho_{0}r_s^3}{\omega}\Big(\frac{r}{r_s}\Big)^{\omega}~ _2F_1[\omega,\omega;1+\omega;-\frac{r}{r_s}]
\end{equation}
where $\omega = 3 - \gamma$.
The final profile we use is a cored generalised NFW (cgNFW)

\begin{equation}
 \rho_{DM}(r) = \rho_{0}\Big(\frac{r+r_c}{r_s}\Big)^{-\gamma}~\Big(1+\frac{r}{r_s}\Big)^{\gamma-3}.
\end{equation}
in which $r_c$ is now the scale radius of the core. This is a more general case of the gNFW, and, in addition to allowing the data to choose between cusps ($r_c = 0$, $\gamma = 1$) and cores ($r_c > 0$, or $r_c = 0$ and $\gamma = 0$), it has additional flexibility at intermediate radii. We carry out the integration of the cumulative mass for this profile numerically.

The BH is simply a point mass at the origin, 

\begin{equation}
 \rho_{BH}(r) = \frac{M_{BH}}{4\pi r^2}\delta(r),
\end{equation}
giving a constant term in the cumulative mass distribution.

As the SAURON dataset far outweighs the GC, NIFS and satellite datasets in terms of size, we regularise its contribution to the likelihood calculation by additionally fitting for a `noise' parameter $\Delta_{\sigma_{\star}^2}$. This can be interpreted as accounting for scatter in the data not included in the uncertainties or, alternatively, as modifying the relative weight given to these data, and is added in quadrature to the measured uncertainties on the velocity dispersion. 

Our overall model therefore has a number of free parameters dependent on the halo model in question and our assumptions about the anisotropy. In the isotropic case, which we treat first, the number of parameters varies between five and seven. In common for all halo models are the mass-to-light ratio $\Upsilon_{\star}$, the normalisation of the DM halo $\log(\rho_0)$, the scale radius $r_s$, the BH mass $\log(M_{BH})$ and the noise in the SAURON data, $\Delta_{\sigma_{\star}^2}$. The gNFW and cgNFW halo models then add the free parameters ($\gamma$), ($\gamma$, $r_c$) respectively. When the orbital anisotropy of each tracer population is allowed to vary, this adds three parameters, as will be discussed in Section~\ref{sec:bayes}. In our notation in the following sections, we use the gNFW parameter set whenever we write out the model parameters explicitly, but this should be understood as standing in for any of the halo models.

\subsection{Jeans modelling}
\label{sec:bayes}

Given the stellar velocity dispersion and the GC and satellite velocities, we want to infer the posterior probability distribution on M87's density profile. For the satellite galaxies, this involves a direct comparison of the mass calculated from our virial estimator and that obtained by integrating the proposed density profile. For the stars and GCs, on the other hand, we can relate the observed velocities and velocity dispersions to the density profile via the Jeans equation. Assuming spherical symmetry and dynamical equilibrium, the Jeans equation has the simple form

\begin{equation}
 \frac{d}{dr}(l\sigma_r^2) + 2\frac{\beta(r)}{r}l\sigma_r^2 = l(r) \frac{GM(r)}{r^2}
\end{equation}
where $l(r)$ is the luminosity density of the tracer, $\sigma_r(r)$ the radial velocity dispersion and $\beta(r)$ is the anisotropy parameter defined in Equation~\ref{eq:binney}. This is a first-order differential equation with general solution

\begin{equation}
 l(r)\sigma_r^2(r) = \frac{1}{f(r)} \int_r^{\infty} f(s) l(s) \frac{G M(s)}{s^2}ds
\end{equation}
where 

\begin{equation}
 f(r) = f(r_i) \exp\Big[\int_{r_i}^r 2 \beta(s) \frac{ds}{s}\Big]
\label{eq:IF}
\end{equation}
\citep[e.g.][]{vanderMarel1994, Mamon2005}. Projecting this along the line of sight gives $\sigma_{los}^2(R)$ as

\begin{equation}
 \frac{1}{2G} I(R) \sigma_{los}(R)^2 = \int_R^{\infty} \frac{l \sigma_r^2 r dr}{\sqrt{r^2 - R^2}} - R^2 \int_R^{\infty}\frac{\beta l \sigma_r^2 dr}{r\sqrt{r^2-R^2}}
\label{eq:jeans}
\end{equation}
which, for certain choices of anisotropy parameterisations, can be written in the simple form

\begin{equation}
 I(R) \sigma_{los}^2(R) = 2G \int_R^{\infty} l K_{\beta} M\frac{dr}{r}
\label{eq:kernel}
\end{equation}
with the kernel $K_{\beta}$ dependent on the particular anisotropy model. Initially, we assume all orbits to be isotropic; later we consider the effect of anisotropy on our inference by modelling the stellar population as following a radially-dependent anistropy profile and each GC population with a constant, non-zero anisotropy. For the stars, we use a scaled Osipkov-Merritt \citep{Osipkov1979, Merritt1985} profile

\begin{equation}
 \beta(r) = \beta_{\infty} \frac{r^2}{r^2 + r_a^2};
\label{eq:scaledOM}
\end{equation}

which is centrally isotropic and becomes radially anisotropic at large radii, tending to the asymptotic anisotropy $\beta_{\infty}$. The kernel for the scaled Osipkov-Merritt profile is presented in the Appendix; the kernel for the constant-anisotropy case is given in \cite{Mamon2005}, and we refer the reader there for further details.

Given a prescription for the anisotropy, the surface brightness and luminosity profiles of Figure~\ref{fig:SBs} and the mass model, then, we can calculate line-of-sight velocity dispersions $\sigma_{los}$ from the Jeans equation and so infer the posterior probability distribution of the model parameters, based on the data in hand.

\subsection{Statistical analysis}
%\label{sub:bayes}

We use the methods of Bayesian analysis to infer the posterior probability distribution of our density profile parameters, given the data. Bayes theorem states that the posterior distribution is proportional to the product of the likelihood function of the data given the model and the priors on the model, and so the task here is to construct sensible likelihood functions for each dataset. The total likelihood is then, in turn, the product of these, as each consitutes an independent set of measurements. 

First, as we have velocity dispersion measurements for the stars, the likelihood of observing a particular velocity dispersion at radius $R$ is assumed to be Gaussian, with a standard deviation equal to the uncertainty. Thus the $k^{th}$ stellar velocity dispersion measurement gives a contribution to the likelihood:

\begin{equation}
 \ln L_{\star,k}(\sigma_k,R_k|\vec{M}) = -0.5\Bigg(\frac{\sigma_k^2 - \sigma_m^2}{\delta_{\sigma_k^2}}\Bigg)^2 - 0.5\ln(2\pi\delta_{\sigma_k^2}^2)
\end{equation}
for uncertainty $\delta_{\sigma_k^2}$ and model prediction $\sigma_m^2$, and model parameters $\vec{M} = (\Upsilon_{\star},M_{BH},\Delta_{\sigma_{\star}^2},\rho_{0},\gamma,r_s)$. As the observations in each aperture are independent, the total log likelihood of observing the ensemble is just the sum:

\begin{equation}
\begin{split}
\ln L_{\star} &= \sum_k \ln L_{\star,k}.
\end{split}
\end{equation}
Note that, for measurement uncertainty $\Delta_{\sigma_k^2}^2$, the regularisation of the SAURON data gives a total uncertainty $\delta_{\sigma_k^2}^2 = \Delta_{\sigma_{k}^2}^2 + \Delta_{\sigma_{\star}^2}^2$. For the NIFS data, on the other hand, $\delta_{\sigma_k^2}^2 = \Delta_{\sigma_{k}^2}^2$.

The virial mass estimate from the satellites can be treated in a similar way, though here we only have one measurement:

\begin{equation}
 \ln L_{sat} = -0.5\Bigg(\frac{\log(M_{sat})-\log(M_{mod})}{\delta M_{sat}}\Bigg)^2 - 0.5 \ln\Big(2\pi\delta M_{sat}^2\Big)
\end{equation}
for virial mass estimate $M_{sat}$, model mass $M_{mod}$ and the logarithm of the uncertainty from the mass estimator, $\delta M_{sat}$. This is a direct comparison without the need for recourse to the Jeans equation. 

For the GCs we can assign probabilities, based on the colour and position information provided in the photometric catalogue, of each GC belonging to either the red or the blue population, although we are not able to classify them with certainty. As the two populations are assumed to be dynamically decoupled, we do not expect their velocity dispersion profiles to be the same. In contrast to the other tracer populations, then, we calculate the likelihood of observing a GC with a particular velocity under the assumption that the velocity distribution of each GC population can be described by a Gaussian with a standard deviation given by the velocity dispersion, such that

\begin{equation}
 \ln L_{GC,k} = -0.5\Bigg(\frac{(v_k-v_{gal})^2}{\delta v_k^2 +\sigma_{m}^2}\Bigg) - 0.5\ln\Big(2\pi(\delta v_k^2 + \sigma_{m}^2)\Big)
\end{equation}
where $v_k$ and $\delta v_k$ are the measured line-of-sight velocity and velocity uncertainty of the $k^{th}$ GC, $v_{gal} = 1284$ kms$^{-1}$ is the heliocentric velocity of M87 \citep{Cappellari2011} and the line-of-sight velocity dispersion at the location of each globular cluster, $\sigma_m$, is modelled separately for the red and the blue populations.

To account for the uncertainty in assigning each GC to either the red or the blue population, for each set of model parameters $\vec{M}$ we draw 1000 Monte Carlo samples which stochastically explore the population distribution based on the colour, magnitude and position information for the individual GCs. We then marginalise over the samples to give a final contribution to the likelihood. While some GCs have either very high or very low probabilities of belonging to one of the GC populations, with small uncertainty, there also exists a significant fraction with comparable probabilities of belonging to either: for these objects, it would not be meaningful to simply assign them to one population or the other. A further advantage of this stochastic sampling is that it allows us to explore different combinations of red and blue GCs.

As each tracer population is independent, the final log-likelihood of any set of model parameters is the sum over all contributions:

\begin{equation}
 \ln L = \sum_k\ln L_{\star,k} + \sum_k\ln L_{GC,k} + \ln L_{M_{sat}}.
\end{equation}
We explore the parameter space using the ensemble-sampling code \texttt{emcee} \citep{Foreman-Mackey2013}.

\section{Results}
\label{sec:results}

\begin{table*}
 \centering
  \begin{tabular}{C{1cm}C{1.4cm}C{1.4cm}cccccc}\hline
  halo model & $M_{\star}/L$ & $\log(\rho_{DM})$ & $r_s$ & $\gamma$ & $\delta_{\sigma_{\star}^2}$ & $r_c$ & $\log(M_{vir})$ & $R_{vir}$ \\\hline
NFW & $6.6 \pm 0.1$ & $6.6 \pm 0.1$ & $448 \pm 75$ & -- & $12.1 \pm 0.2$ & -- & $14.39_{-0.53}^{+1.11}$ & $1620_{-360}^{+460}$ \\
  LOG & $6.9 \pm 0.1$ & $7.7 \pm 0.1$ & $48 \pm 5$ & --  & $12.0 \pm 0.2$ & -- & $14.21_{-0.37}^{+0.60}$   & $1410_{-200}^{+240}$  \\
  gNFW & $6.9 \pm 0.1$ & $8.3 \pm 0.1$ & $79 \pm 10$ & $ < 0.14$ & $12 \pm 0.1$ & -- & $14.13_{-0.44}^{+0.76}$   & $1320_{-240}^{+270}$  \\
  cgNFW* & $6.9 \pm 0.1$ & $5.6_{-1.2}^{+1.4}$ & $273_{-180}^{+430} $ & $2.54_{-1.18}^{+0.33}$ & $12.0 \pm 0.2$ & $63_{-14}^{+11}$ & $14.16_{-0.40}^{+0.67}$ & $1350_{-200}^{+240}$  \\\hline
  \end{tabular}
\caption{Final inference on the parameters in the different $\beta=0$ models. We report the median values of our inferred posterior distributions, along with the 16th and 84th percentiles as a measure of our uncertainty. For the inner slope $\gamma$ in the gNFW and cgNFW models, the 95 \% confidence value is given. All quantities are measured in units of solar mass, solar luminosity, kilometers per second and kiloparsecs. *Note that the cgNFW posterior is bimodal due to degeneracies inherent in the profile: see the panel of Figure~\ref{fig:triangleBBPL}.}
\label{tab:results}
\end{table*}

\begin{figure*}
  \centering
  \includegraphics[width=1\textwidth]{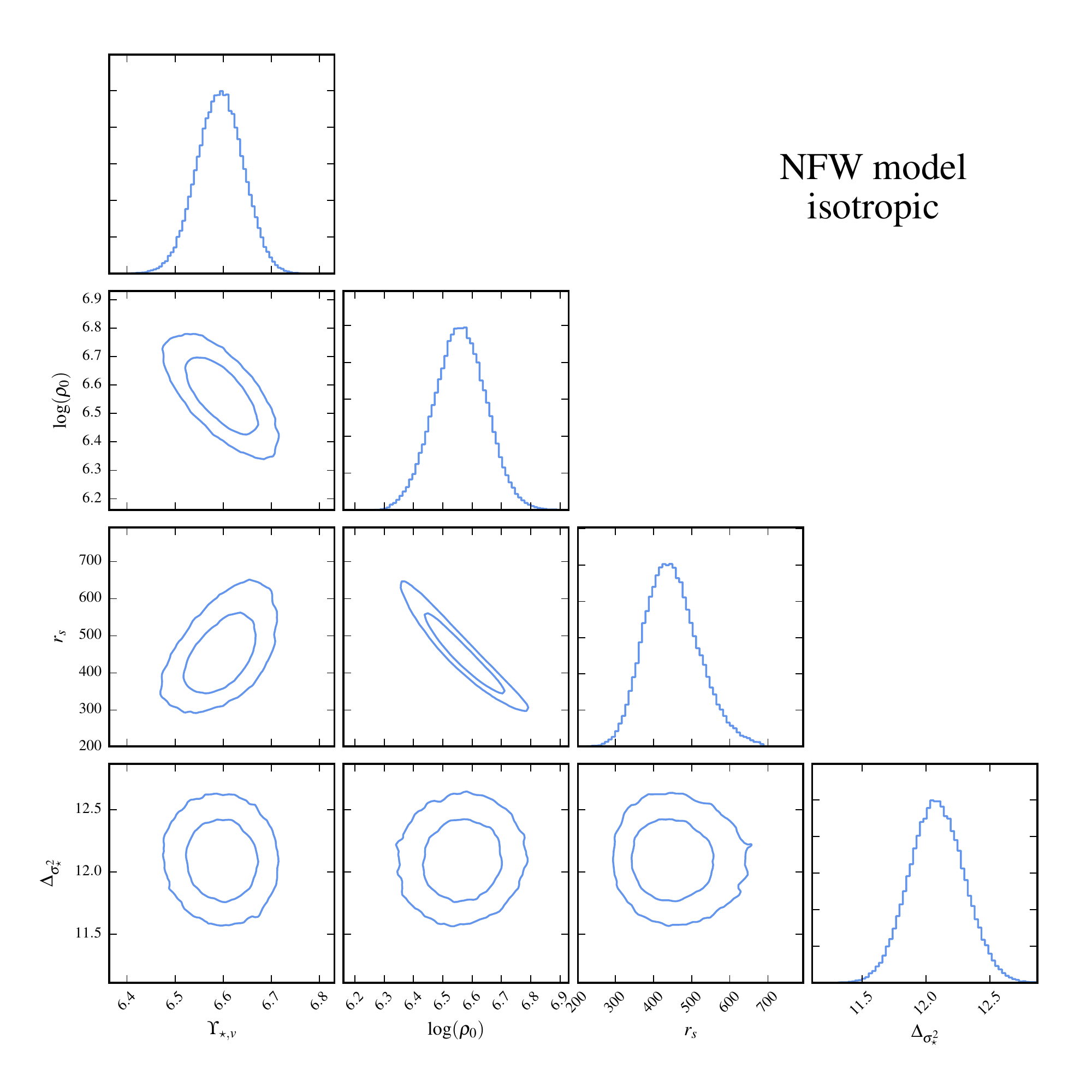}
  \caption{Inference on the NFW model parameters, assuming isotropy. This model favours a lower mass-to-light ratio than the others, due to the larger amount of DM that the cusp necessarily puts at small radii. As in Tables 2 and 3, all quantities are measured in units of solar mass, solar luminosity, kilometers per second and kiloparsecs.}
\label{fig:triangleNFW}
\end{figure*}

\begin{figure*}
  \centering
  \includegraphics[width=1\textwidth]{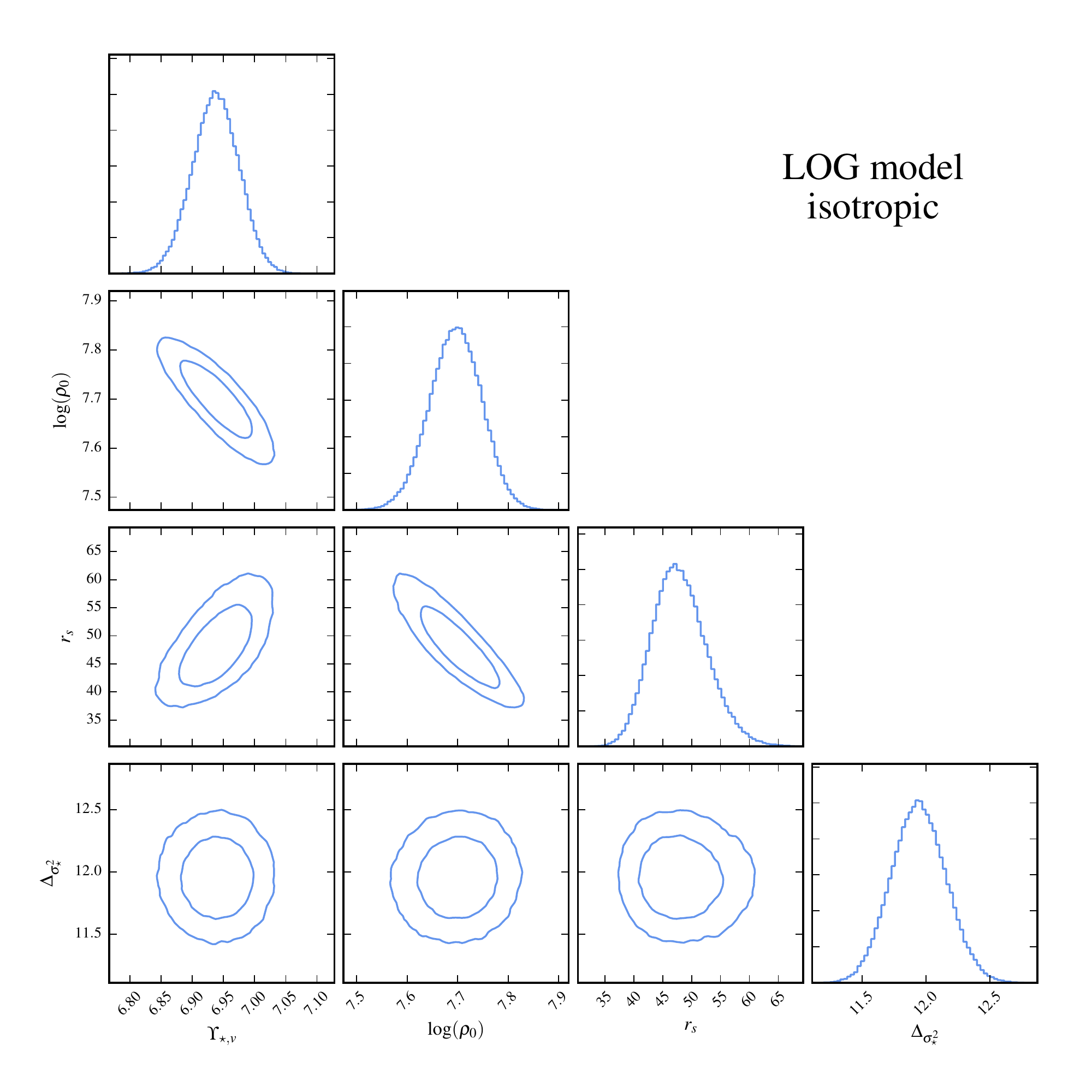} 
  \caption{Inference on the LOG model parameters, assuming isotropy. All quantities are measured in units of solar mass, solar luminosity, kilometers per second and kiloparsecs.}
\label{fig:triangleLOG}
\end{figure*}

\begin{figure*}
  \centering
  \includegraphics[width=1\textwidth]{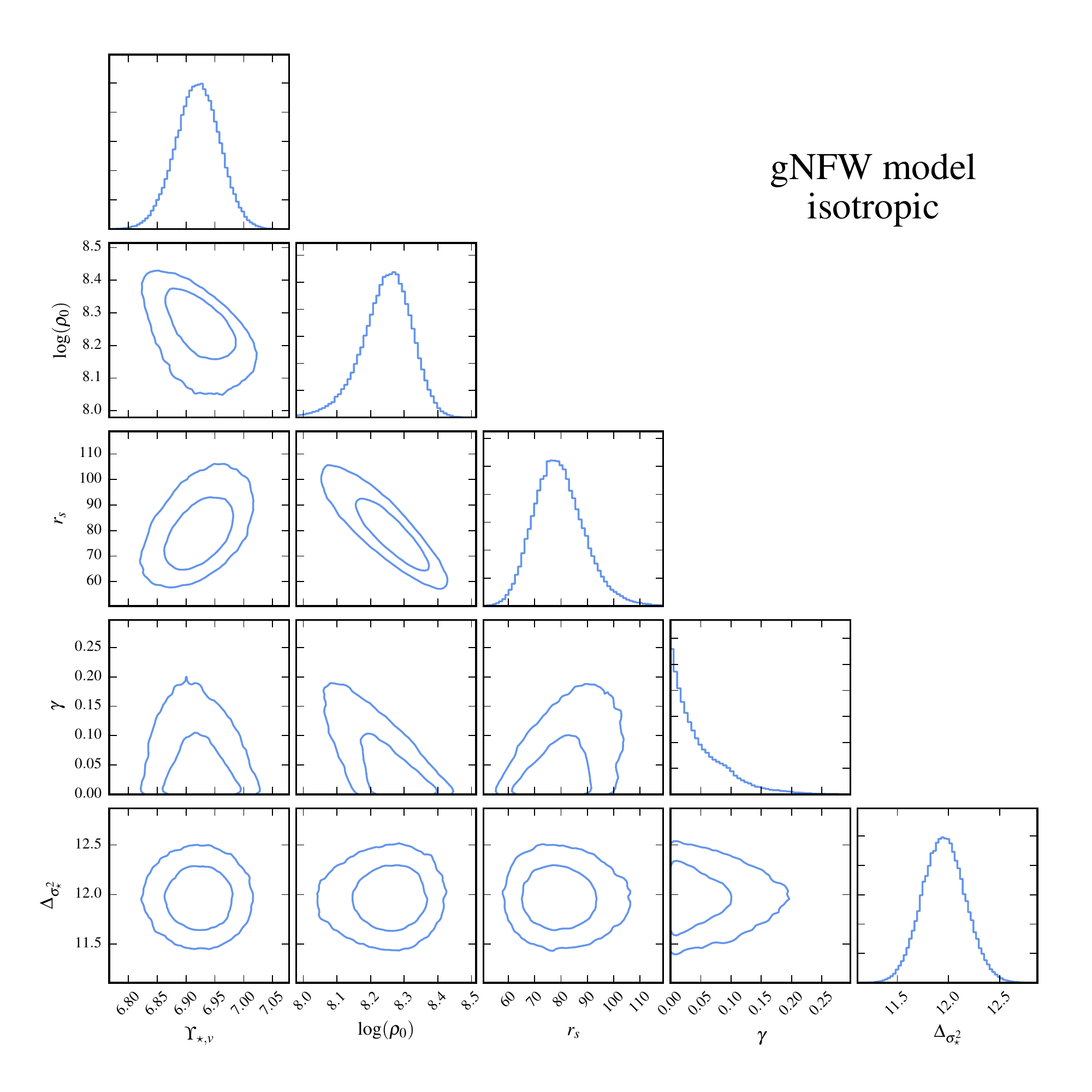} % use iCornerPlotter from within the terminal to get the right font!
  \caption{Inference on the gNFW model parameters, assuming isotropy. Note that the posterior on the inner slope, $\gamma$, hits the lower limit imposed by the prior. All quantities are measured in units of solar mass, solar luminosity, kilometers per second and kiloparsecs.}
\label{fig:triangleBPL}
\end{figure*}

\begin{figure*}
  \centering
  \includegraphics[width=1\textwidth]{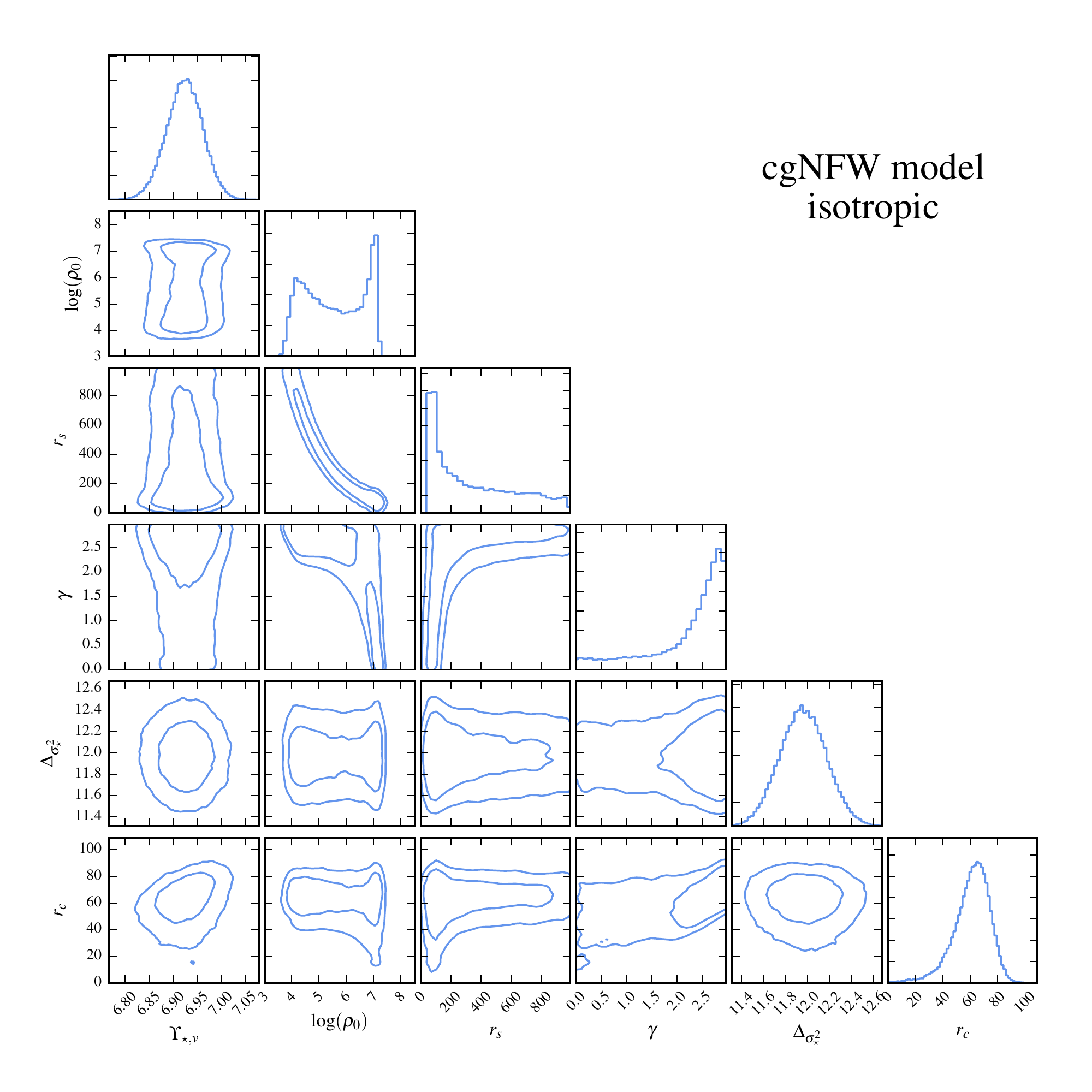} 
  \caption{Inference on the cgNFW model parameters, assuming isotropy. All quantities are measured in units of solar mass, solar luminosity, kilometers per second and kiloparsecs. Note that our cgNFW modelling implies that the halo is consistent with a cored power law with power law index 3. For the cgNFW profile, this leads to a degeneracy in the parameters $\gamma$, $r_s$ and $r_c$: given a finite core radius, the same solution can be obtained either by having $\gamma \sim 0$ and $r_s \sim r_c$ (in which case $\rho_0$ is large), or by having $\gamma \sim 3$, in which case $r_s$ (and so $\rho_0$) is relatively unconstrained. This is what we find here, producing the apparent bimodality in $\rho_0$. Both these solutions correspond to the same mass profile.}
\label{fig:triangleBBPL}
\end{figure*}

\begin{figure*}
  \centering
  \subfigure{\includegraphics[trim=15 0 5 0,clip,width=0.5\textwidth]{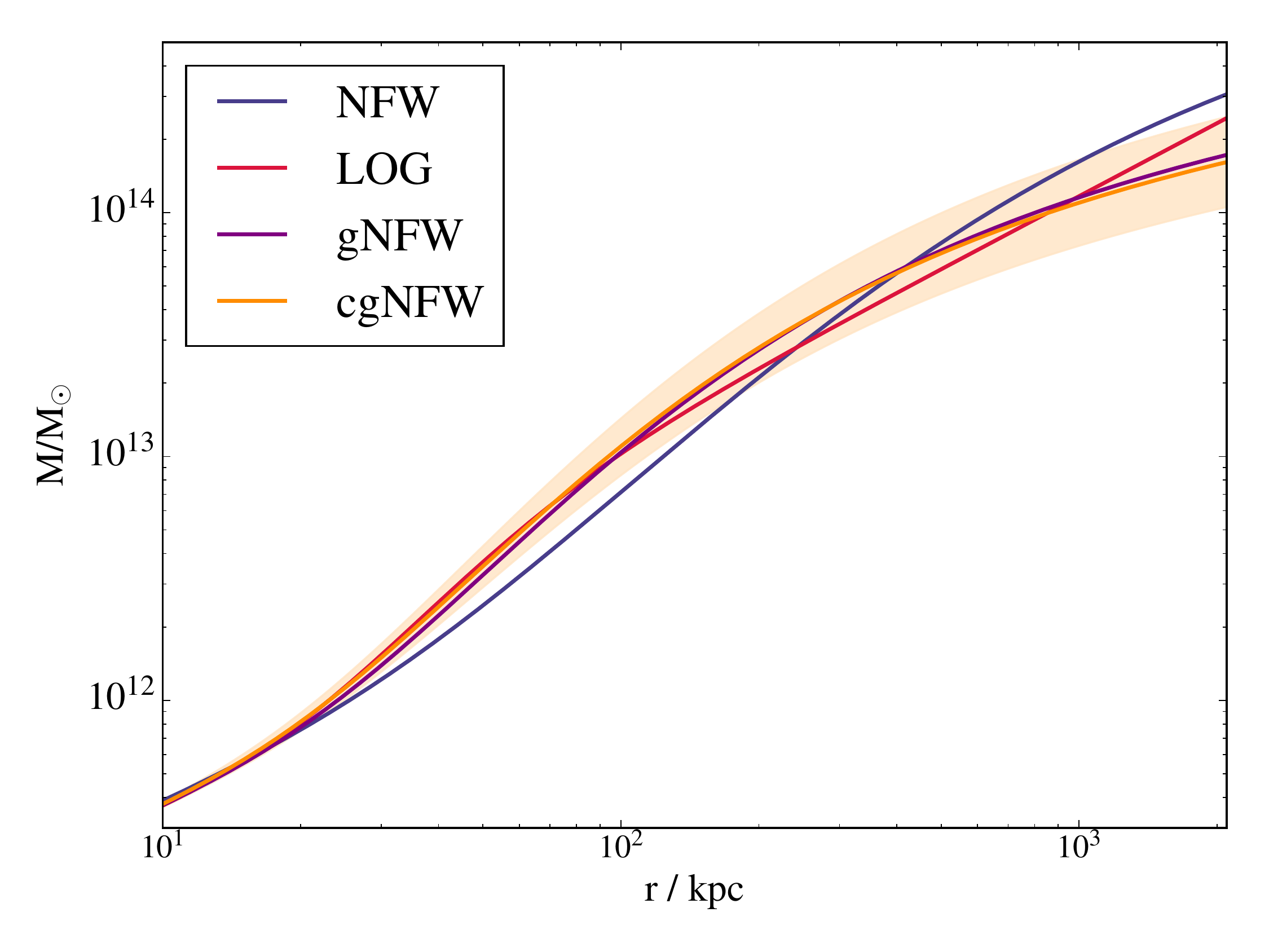}}\hfill
  \subfigure{\includegraphics[trim=15 0 5 0,clip,width=0.5\textwidth]{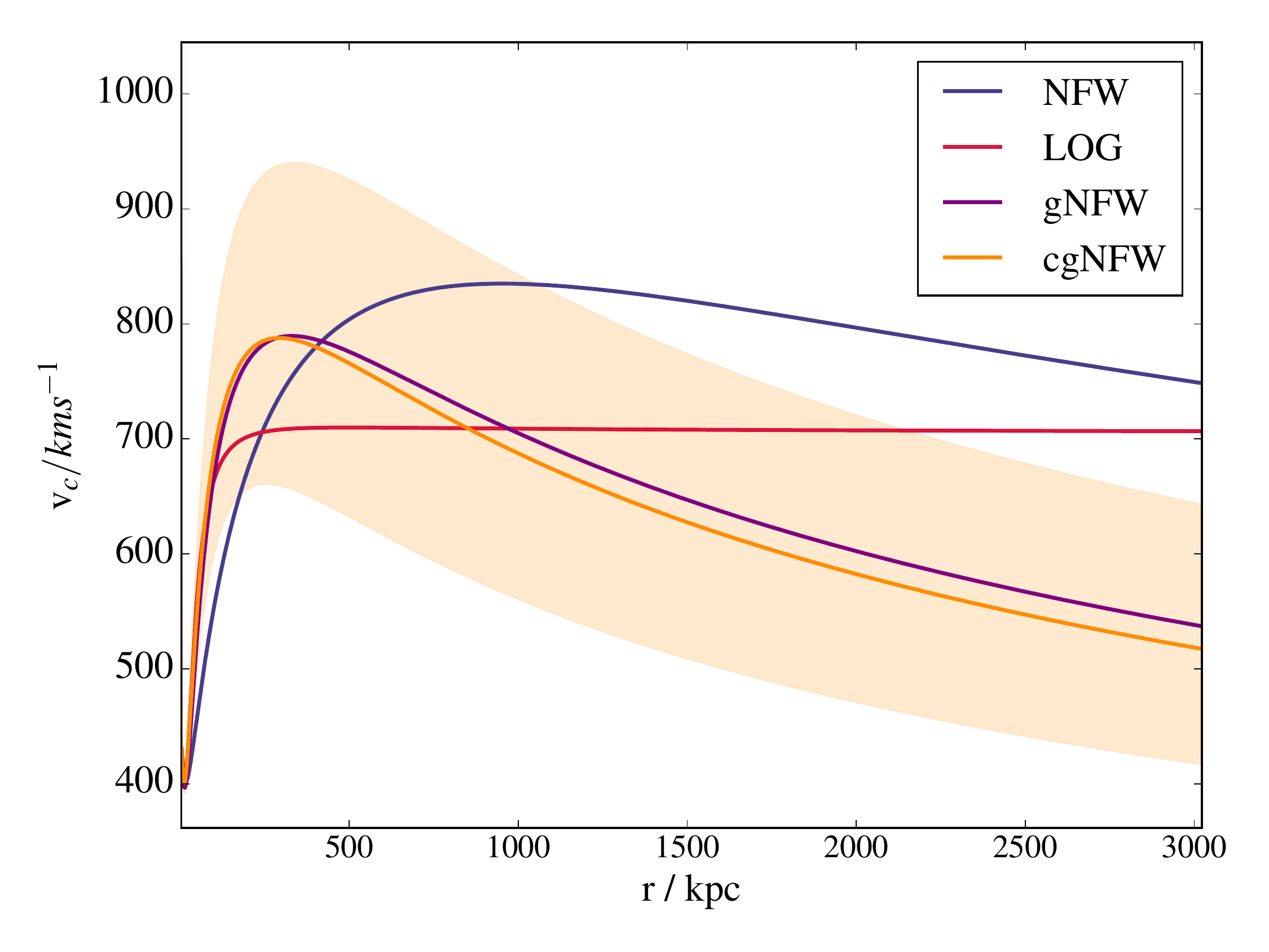}}\hfill % or vcirc.pdf for a linear x-axis.
 \caption{Left: mass profiles inferred using the four different halo models, under assumptions of isotropy. At small radii, all are dominated by the stellar mass, so there is little scatter. At larger radii they differ due to the different structures imposed by the halo models: the LOG density profile, for instance, decays as $r^{-2}$ while the others go as $r^{-3}$, and this places more mass further out in the halo. The NFW is the only profile which does not allow a central core, and its mass profile differs from the other three across a wide radius range. Right: circular velocity curves for the four halo models. This highlights the differences between them. In both figures, the band indicates the 68 \% uncertainty region for the cgNFW model.}
 \label{fig:MassProfiles}
\end{figure*}

\begin{figure*}
  \centering
  \includegraphics[width=1\textwidth]{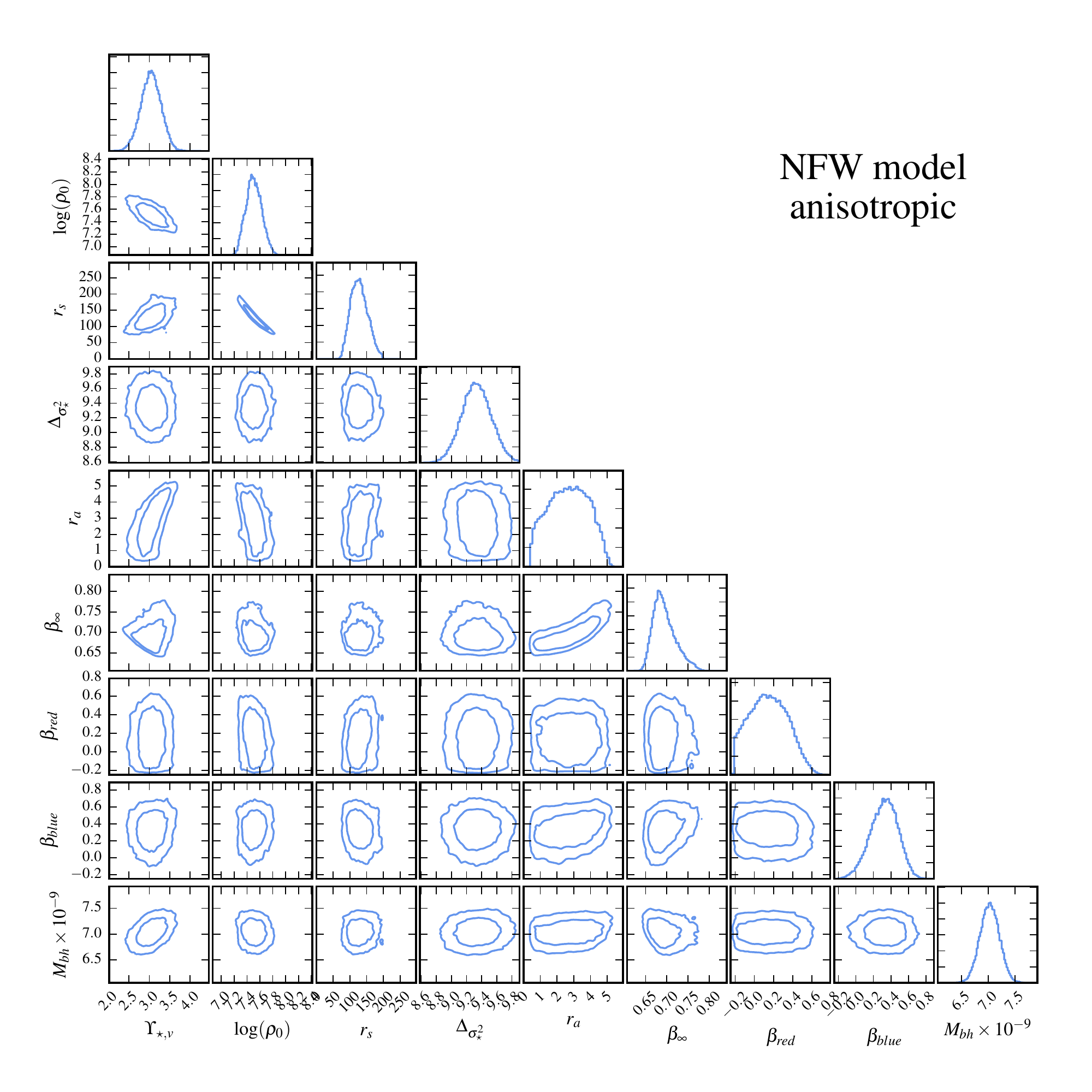} 
  \caption{Inference on the NFW model parameters, allowing for anisotropy. We characterise the stars with a scaled Osipkov-Merritt profile which becomes isotropic centrally and tends to $\beta = \beta_{\infty}$ at large radii; for simplicity, each GC population is modelled with constant anisotropy. All quantities are measured in units of solar mass, solar luminosity, kilometers per second and kiloparsecs.}
\label{fig:triangleNFWaniso}
\end{figure*}

\begin{figure*}
  \centering
  \includegraphics[width=1\textwidth]{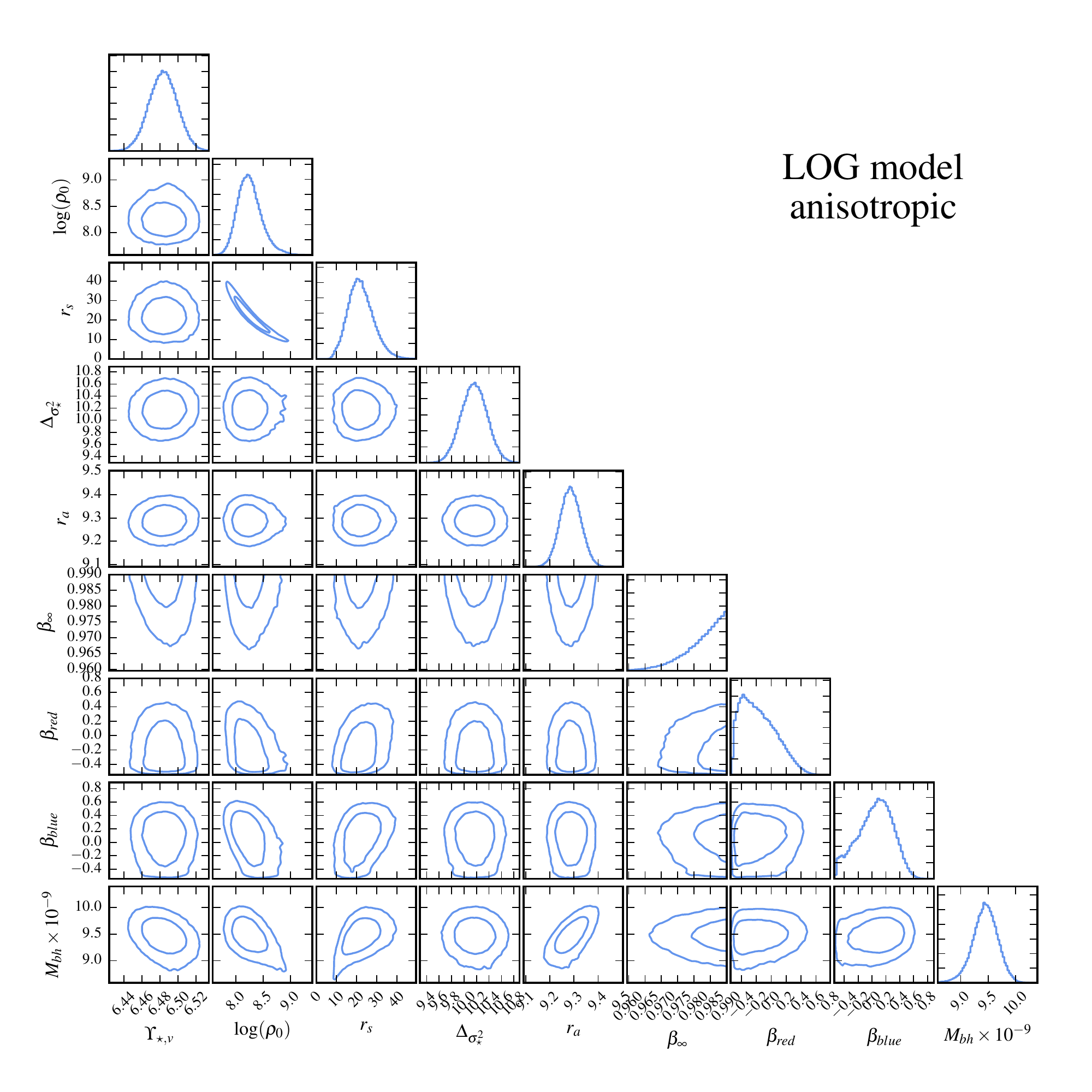} 
  \caption{Inference on the LOG model parameters, allowing for anisotropy. We characterise the stars with a scaled Osipkov-Merritt profile which becomes isotropic centrally and tends to $\beta = \beta_{\infty}$ at large radii; for simplicity, each GC population is modelled with constant anisotropy. All quantities are measured in units of solar mass, solar luminosity, kilometers per second and kiloparsecs.}
\label{fig:triangleLOGaniso}
\end{figure*}

\begin{figure*}
  \centering
  \includegraphics[width=1\textwidth]{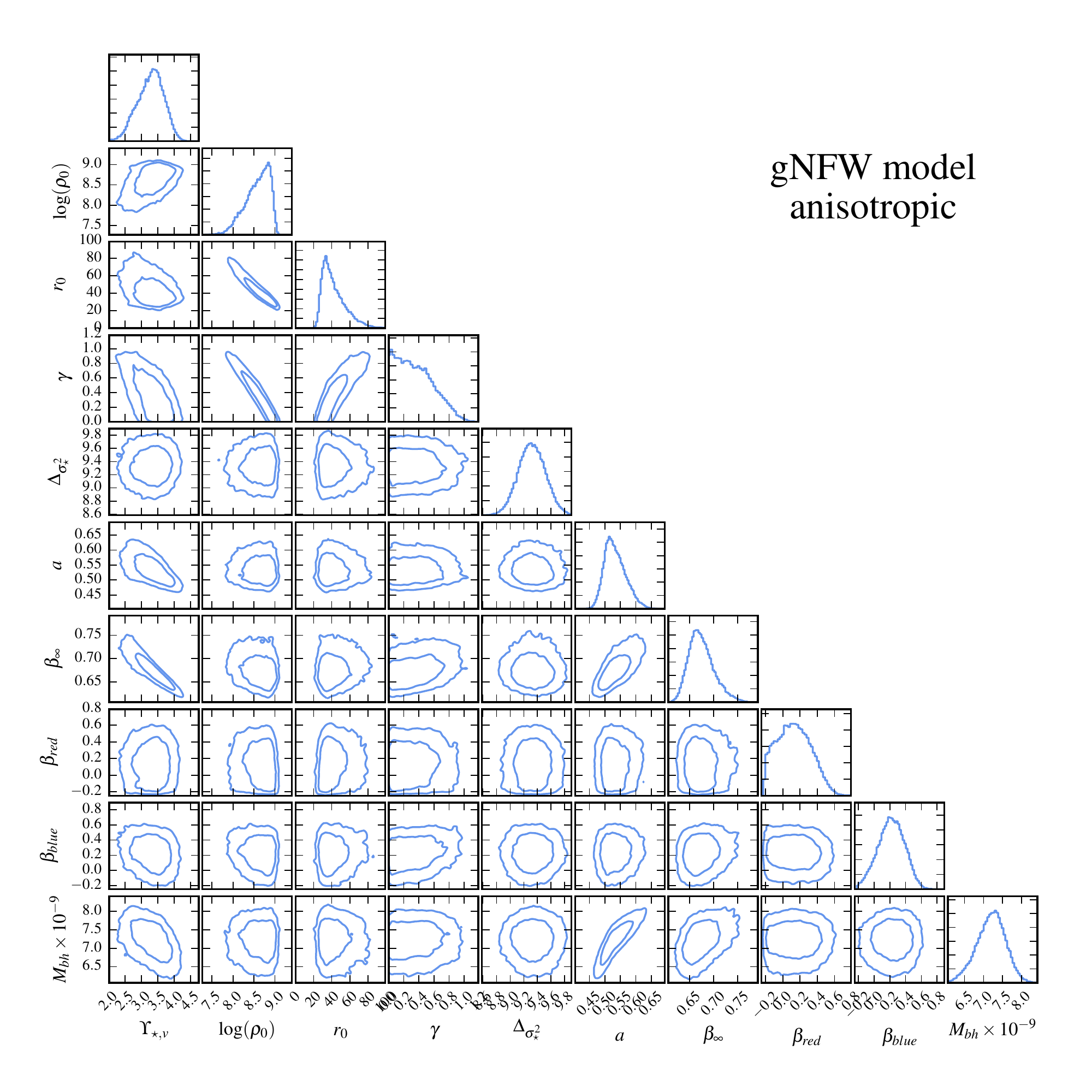} 
  \caption{Inference on the gNFW model parameters, allowing for anisotropy. We characterise the stars with a scaled Osipkov-Merritt profile which becomes isotropic centrally and tends to $\beta = \beta_{\infty}$ at large radii; for simplicity, each GC population is modelled with constant anisotropy. All quantities are measured in units of solar mass, solar luminosity, kilometers per second and kiloparsecs.}
\label{fig:triangleBPLaniso}
\end{figure*}

\begin{figure*}
  \centering
  \includegraphics[width=1\textwidth]{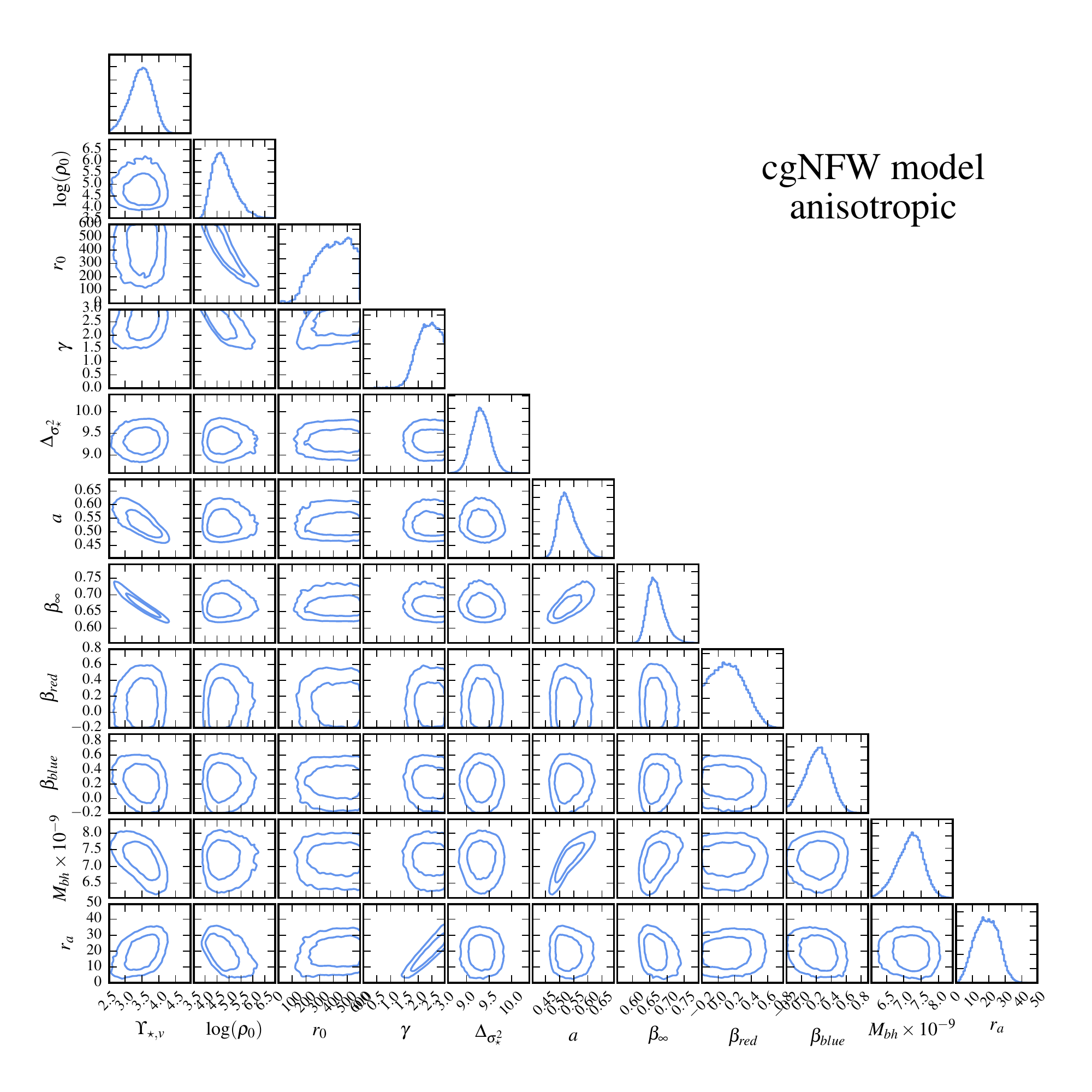} 
  \caption{Inference on the cgNFW model parameters, allowing for anisotropy. We characterise the stars with a scaled Osipkov-Merritt profile which becomes isotropic centrally and tends to $\beta = \beta_{\infty}$ at large radii; for simplicity, each GC population is modelled with constant anisotropy. All quantities are measured in units of solar mass, solar luminosity, kilometers per second and kiloparsecs.}
\label{fig:triangleBBPLaniso}
\end{figure*}

%%%

Initially, we treat all orbits as isotropic and investigate the uncertainty introduced to the mass inference by the choice of halo model. The inference for the four different halo models are presented in Figures~\ref{fig:triangleNFW} - \ref{fig:triangleBBPL}, and are summarised in Table~\ref{tab:results} along with the associated virial parameters. We find the results of the gNFW, cgNFW and LOG models to be very similar: the gNFW and cgNFW haloes are both centrally cored with break radii $\sim$ 80 kpc, while the LOG model is inherently cored and has a slightly smaller scale radius $r_s = 48 \pm 5$ kpc (note, however, that these are \textit{not} core radii, and that the scale radius is defined differently in the latter case). All three models also agree closely on the high stellar mass-to-light ratio $\Upsilon_{\star,v} \sim 6.9$ and the virial parameters, with $\log(M_{vir}/M_{\odot}) \sim 14.2$  and R$_{vir} \sim 1.3$ Mpc. Indeed, the best cgNFW model approximately recovers the gNFW solution, with an intermediate slope $\gamma \sim 3$ and both the core and break radii comparable to the gNFW scale radius. Like the gNFW, this represents a solution in which the halo is cored centrally and becomes NFW-like at large radii; unlike the gNFW, it allows more flexibility in the intermediate regions, and the fact that this intermediate slope is found to be close to the NFW value $\gamma \sim 3$ suggests that the NFW profile is an adequate description at intermediate radii as well as large radii. 

On the other hand, the NFW model predicts a lower mass-to-light ratio $\Upsilon_{\star,V} = 6.60 \pm 0.05$ , though it does match the other models in terms of its virial parameters. The difference at small radii arises because of the NFW profile's hard-wired cusp, which places more DM in the central regions at the expense of stellar mass. In Figure~\ref{fig:MassProfiles} we plot the mass profiles inferred in each case, along with the associated rotation curves in order to better highlight the differences between the four models. As can be seen, deviations start to arise at larger radii, where the constraints from our data are weakest. These are mainly due to the different structural features of the models: for instance, the LOG profile decays as $r^{-2}$ at large radii whereas the other three go as $r^{-3}$, which allows the former to place more mass at large radii. Equally, the NFW profile goes as $r^{-1}$ at small radii whereas the others are either intrinsically cored or have flexible inner slopes (which are found to favour cores here), and this causes the NFW profile to start deviating from the others at smaller radii. 

Following this, we test the robustness of our inference on the halo structure against model assumptions by relaxing the isotropy condition and introducing some element of anisotropy into each of the tracer populations. Guided by cosmological N-body simulations, \citep[e.g.][]{Diaferio1999,  Colin2000, Wojtak2005}, which find orbits to change smoothly from being radially anisotropic in the galaxy outskirts to being isotropic in the centre -- a result of hierarchical formation in which BCGs form by the accretion of infalling satellites which become phase-mixed by the time they reach the galaxy centre -- we choose to model the stellar anisotropy using the scaled Osipkov-Merritt profile of Equation~\ref{eq:scaledOM}. As the GC data are comparatively sparse, we treat each GC population as having some constant, non-zero anisotropy. This increases our parameter space by four, as we are now also fitting for the scale radius of the stellar scaled Osipkov-Merrit profile $r_a$, the asymptotic stellar anisotropy $\beta_{\infty}$, and the anisotropies of the two GC populatons, $\beta_{red}$ and $\beta_{blue}$. Again, we carry out the inference for the four halo models, and present the posteriors in Figures~\ref{fig:triangleNFWaniso} - \ref{fig:triangleBBPLaniso} and summarise the results in Table~\ref{tab:anisoresults}. 

Interestingly, we now find the some significant deviation among the halo models. Firstly, the LOG profile recovers a model for the stars that is quite isotropic, with $\beta_{\infty} > 0.99$ at the 95\% confidence level and $r_a = 9.29\pm 0.05$ kpc; given that the SAURON data only extend out to $\sim 2$ kpc, this implies anisotropies $\beta < 0.05$ where we have data (though our data are in projection). This is presumably a result of the hard-wired core. The remaining three models agree, within uncertainties, on the mass profile -- though, as might be expected, not so closely as when isotropy is enforced. In these cases, we find that the mass-to-light ratio drops significantly and that the stars become radially anisotropic within scale radii $r_a < 4$ kpc. This covariance between the central mass and anisotropy is a direct manifestation of the mass-anisotropy degeneracy: in the isotropic models, the assumed lack of any radial anisotropy drove the central mass to high values, whereas here we are able to go some way towards breaking the degeneracy through the use of multiple populations tracing the same gravitational potential. Note also that the introduction of radial anisotropy, coupled with the decrease in stellar mass, now allows the DM halo to become less cored in the centre: we find a gNFW inner slope which is still distinctly sub-NFW, but less extremely so, with $\gamma < 0.81$ at the 95 \% confidence level, and the cgNFW model finds a smaller central core ($r_c = 19.00_{-8.34}^{+8.38}$ kpc) and a sub-NFW intermediate slope $\gamma = 2.39_{-0.43}^{+0.39}$ before transitioning to the $r^{-3}$ regime beyond the scale radius $r_s = 412.1_{-143.0}^{+123.0}$ kpc. For these three profiles, both GC populations are found to have a mild radial anisotropies. We find a slightly higher BH mass than previous estimates \citep[e.g.][]{Gebhardt2011}, though it is consistent within uncertainties. However, we note that this is the first time the BH mass has been inferred jointly with the halo and stellar mass parameters using the high-spatial-resolution NIFS dataset; as we might anticipate a strong covariance between these parameters -- especially between the stellar and BH masses -- a slight change is not too surprising. On the other hand, there are still assumptions in our model which could affect our inference on the BH mass. For instance, our anisotropy profile enforces isotropy in the centre, which, while a physically sound assumption, restricts the range of parameter space explored by the BH mass. We are also only making use of the second-order moments of the line-of-sight velocity profile: while this should not cause any systematic bias in the BH mass inference, we may be able to obtain tighter constraints by incorporating higher-order moments.

\begin{figure}
  \centering
  \subfigure{\includegraphics[trim=15 0 0 0,clip,width=0.5\textwidth]{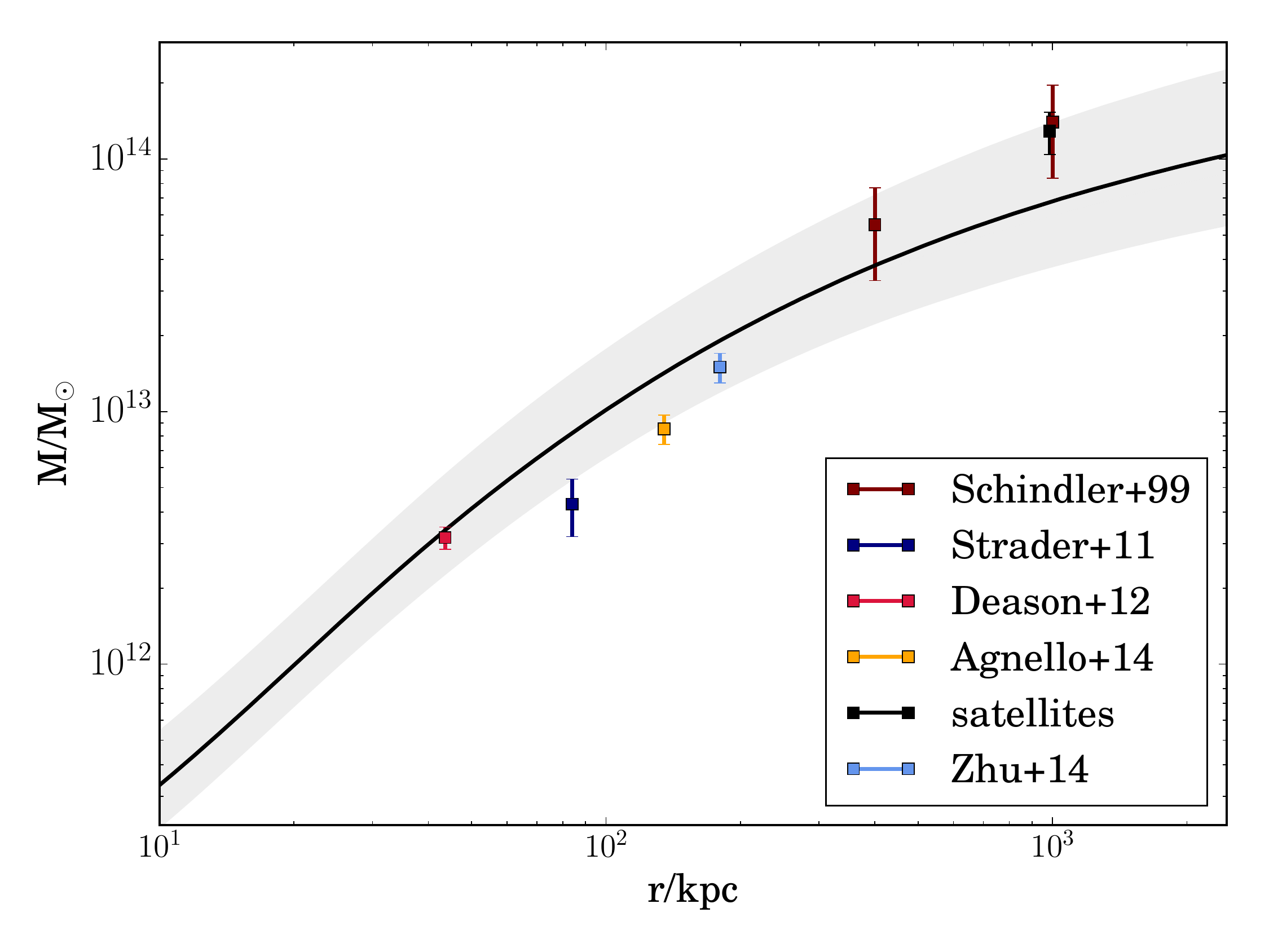}}\hfill 
    \caption{Final inference on the mass profile for the anisotropic cgNFW model, with 1 $\sigma$ uncertainties. Overplotted are the mass measurements reported from other studies: these generally bracket our result well.}
\label{fig:mass}
\end{figure}

\begin{figure}
  \centering
  \subfigure{\includegraphics[trim=15 0 15 0,clip,width=0.52\textwidth]{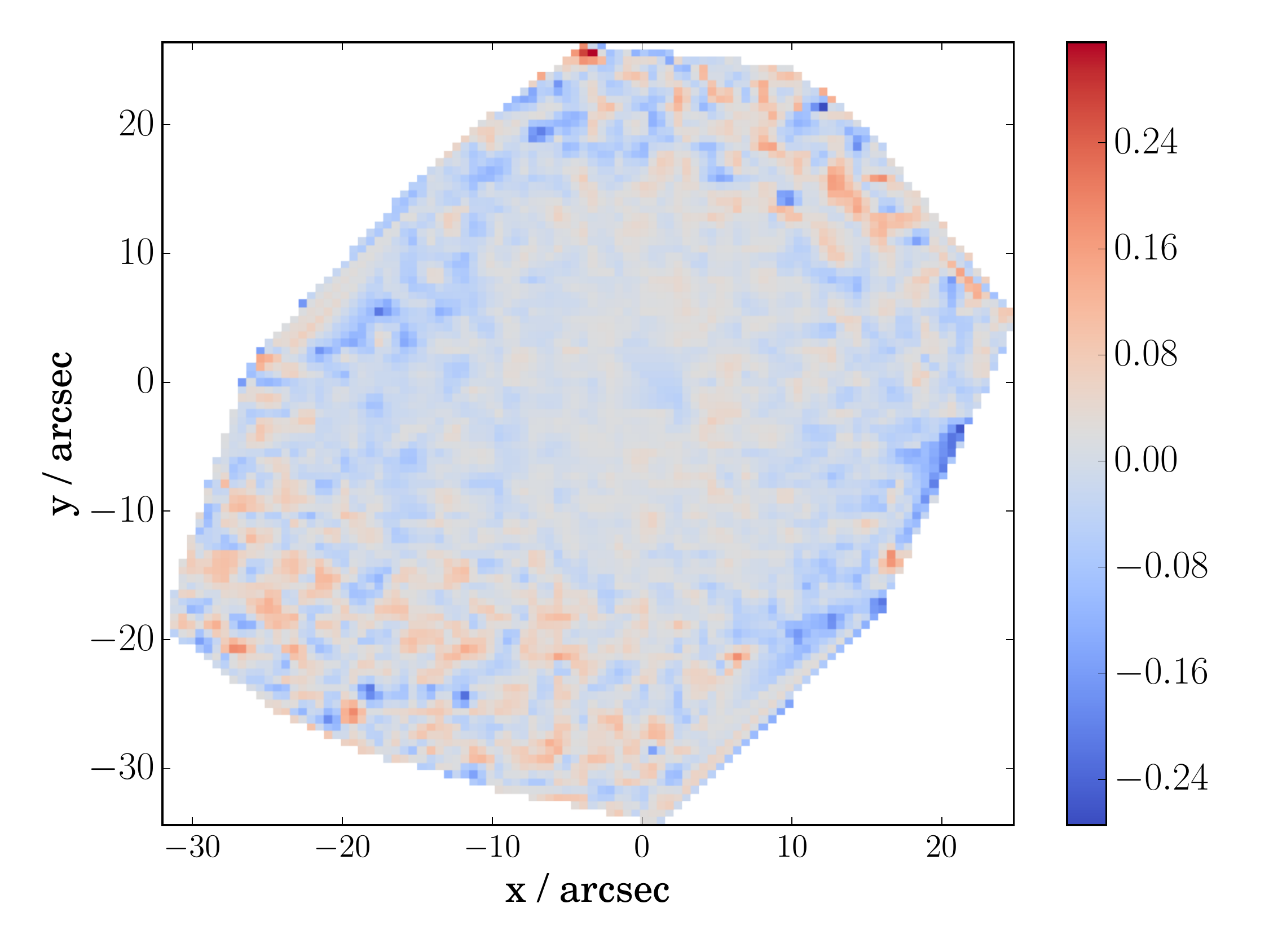}}\hfill % vcirc1.pdf has no error bounds
  \caption{Fractional residuals on the stellar velocity dispersion data for the anisotropic cgNFW model.}
\label{fig:residuals}
\end{figure}

We compare the goodness of fit of the four models using a reduced chi-squared criterion, and find that the cgNFW profile provides the best description of the dynamics -- noting that the cgNFW, gNFW and NFW are nested models. We therefore present the best cgNFW mass profile in Figure~\ref{fig:mass} and the residuals on the SAURON and NIFS data in Figure~\ref{fig:residuals}. The mass profile appears to be generally consistent with earlier work; the residuals are noticeably better than in the isotropic case, indicating the importance of the anisotropy in the stellar dynamics.

We also note that the assumption that the MultiDark simulations are able to accurately reproduce the dynamics of real galaxies may introduce some additional systematic uncertainty into the mass measurement made using the virial estimator in Section~\ref{sec:data:satellites}. We therefore test the sensitivity of our inference to this uncertainty by carrying out the analysis with twice the calculated uncertainty, and we find that this has a negligible effect on our inferred density model. Essentially, the likelihood is dominated by contributions from the stars and the GCs, as the datasets for these populations are far more extensive; the satellite mass estimate rejects models with discrepant extrapolations of the total mass to large radii, but most of the models consistent with the smaller-radius data are still well-behaved at 1 Mpc.

\begin{table*}
\tiny
 \centering
  \begin{tabular}{cC{0.8cm}C{1.0cm}C{1.0cm}C{1.0cm}C{1.0cm}C{1.0cm}C{1.0cm}C{1.0cm}C{1.0cm}C{1.0cm}C{1.0cm}C{1.0cm}C{1.0cm}}\hline
   & $M_{\star}/L$ & $\log(\rho_{DM})$ & $r_s$ & $\gamma$ & $M_{BH} \times 10^{-9}$ & $\delta_{\sigma_{\star}^2}$ & $r_a$ & $\beta_{\infty}$ & $\beta_{red}$ & $\beta_{blue}$ & $r_c$ & $\log(M_{vir})$ & $R_{vir}$ \\\hline
NFW & $3.05_{-0.24}^{+0.24}$ & $7.51_{-0.12}^{+0.13}$ & $128.4_{-24.1}^{+27.8}$ & -- & $7.04_{-0.2}^{+0.2}$ & $9.34_{-0.19}^{+0.20}$ & $2.78_{-1.30}^{+1.26}$ & $0.69_{0.04}^{+0.04}$ & $0.16_{-0.21}^{+0.21}$ & $0.34_{-0.17}^{+0.15}$ & -- & $14.00_{-0.32}^{+0.37}$ & $1190_{-120}^{+240}$ \\
LOG & $6.48_{-0.02}^{+0.02}$ & $8.25_{-0.19}^{+0.23}$ & $22.1_{-5.7}^{+6.7}$ & -- & $9.5_{-0.23}^{+0.22}$ & $10.2_{-0.2}^{+0.2}$ & $9.29_{-0.05}^{+0.05}$ & $>0.99$ & $-0.17_{-0.21}^{+0.23}$ & $0.08_{-0.29}^{+0.23}$ & -- & $14.03_{-0.29}^{+0.44}$ & $1200_{-190}^{+250}$ \\
gNFW & $3.33_{-0.45}^{+0.38}$ & $8.64_{-0.35}^{+0.24}$ & $40.3_{-8.8}^{+16.4}$ & $< 0.81$ & $7.2_{-0.4}^{+0.35}$ & $9.3_{-0.2}^{+0.2}$ & $1.06_{-0.10}^{+0.07}$ & $0.67_{-0.05}^{+0.05}$ & $0.13_{-0.20}^{+0.21}$ &  $0.23_{-0.17}^{+0.17}$ & -- &  $13.74_{-0.32}^{+0.43}$ & $990_{-140}^{+270}$ \\
cgNFW & $3.50_{-0.36}^{+0.32}$ & $4.76_{-0.40}^{+0.55}$ & $412.1_{-143.0}^{+123.0}$ & $2.39_{-0.43}^{+0.39}$ & $7.22_{-0.40}^{+0.34}$ & $9.33_{-0.20}^{+0.21}$ & $1.06_{-0.05}^{+0.08}$ & $0.67_{-0.06}^{+0.03}$ & $0.13_{-0.20}^{+0.21}$ & $0.23_{-0.18}^{+0.16}$ & $19.00_{-8.34}^{+8.38}$ & $13.87_{-0.35}^{+0.42}$ & $1090_{-150}^{+220}$ \\\hline
  \end{tabular}
\caption{Final inference on the parameters for the anisotropic models. We report the maximum-likelihhood values of our inferred posterior distributions, along with the 16th and 84th percentiles as a measure of our uncertainty. All quantities are measured in units of solar mass, solar luminosity, kilometers per second and kiloparsecs.}
\label{tab:anisoresults}
\end{table*}

\section{Discussion}
\label{sec:discussion}

\subsection{Resolving previous discrepancies}

Part of the impetus for this study was to resolve the discrepancies between recent models of M87's mass structure. The two recent analyses of \cite{Zhu2014} and \cite{Agnello2014}, both of which relied heavily on the \cite{Strader2011} GC kinematics, disagreed on both the structure of the DM halo and the total stellar mass of the system, with \cite{Agnello2014} finding a stellar mass-to-light ratio $\Upsilon_{\star,V} \sim 4.5$ and a super-NFW cusp $\gamma = 1.6$ and \cite{Zhu2014} inferring $\Upsilon_{\star,I} = 6.0 \pm 0.3 $ (which converts to a $V$-band measurement of $\sim 7.5$) under the assumption of a cored halo. There are a number of possible reasons for this: first, while both studies overlapped in the majority of their GC data, each used a different mass model, and it is possible that these may have unnecessarily constrained the inference on the mass. Specifically, \cite{Zhu2014} chose to use a logarithmic potential model for the dark matter, so enforcing a core, whereas \cite{Agnello2014} used a power law for the dark halo, thus requiring a constant slope at all radii. Neither of these allows the halo much flexibility in the central regions. Further, while \cite{Agnello2014} relied solely on the GCs, separating them into three independent populations based on their velocities, positions and colours, \cite{Zhu2014} treated all the GCs as a single tracer population, but used the same SAURON data as in this study to constrain the stellar mass-to-light ratio. Thus it is possible that \cite{Agnello2014} lacked the data coverage in these central regions that would have permitted a reliable distinction between cusps and cores.

\begin{figure*}
  \centering
  \includegraphics[width=\textwidth]{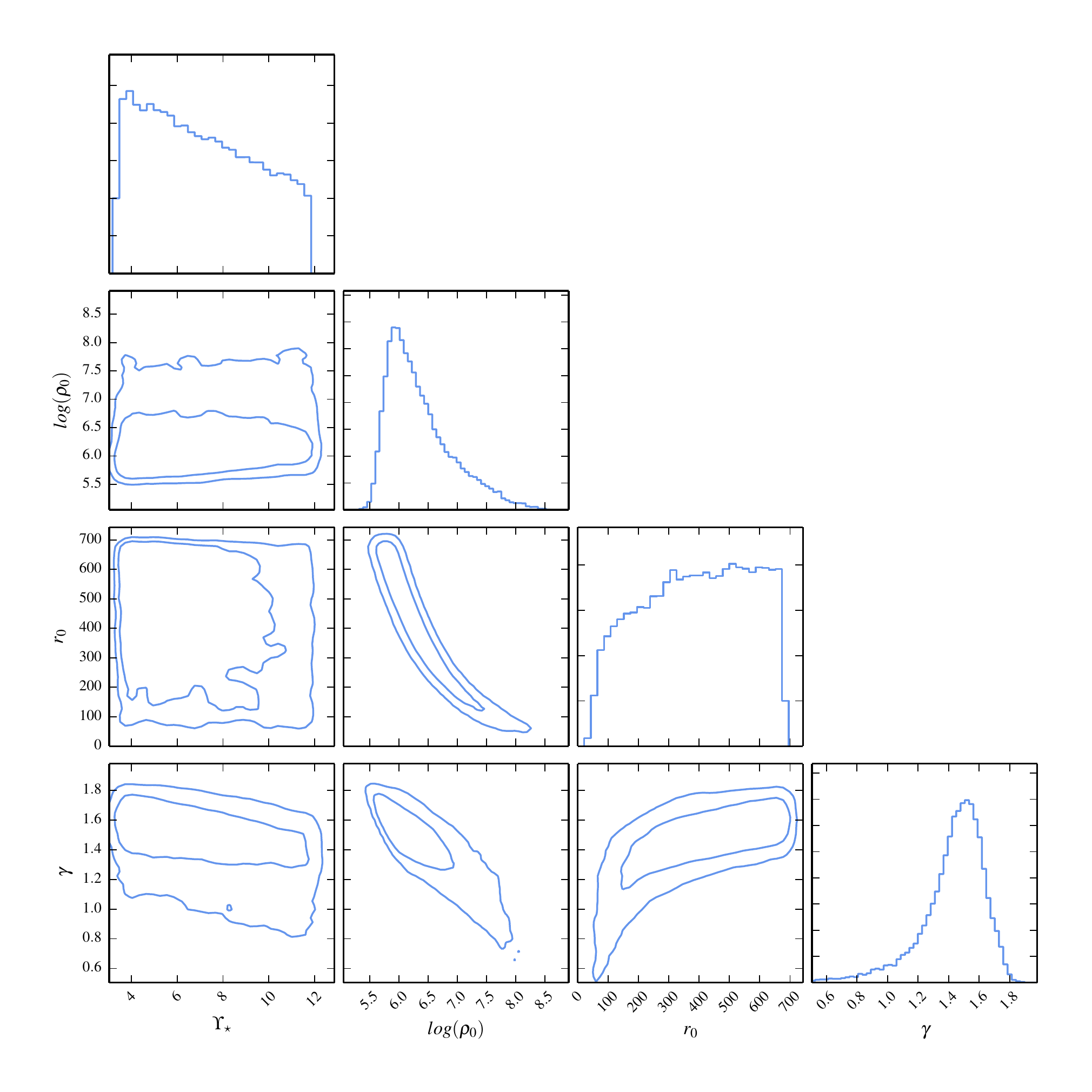}\hfill % using fchaincut_MODEL2_GCsonly.txt
  \caption{Inference on the gNFW model parameters when the stellar data are excluded. The stellar mass-to-light ratio and halo scale radius are unconstrained and the effect of the priors is visible for these parameters, while the halo becomes cuspy, with inner slope $\gamma = 1.48 \pm 0.22$. This is a very different result from the case where all data are modelled in conjunction, and shows the value of the multiple-population method.}
\label{fig:cusp}
\end{figure*}

By exploring multiple mass models and data combinations, we are able to reproduce the results of both studies. First, excluding the stellar kinematics and carrying out the inference using just the satellites and GCs in a way more similar to that of \cite{Agnello2014}, we infer an isotropic gNFW profile with $\gamma = 1.48 \pm 0.22$, thus reproducing their finding of a cusp. The inference is summarised in Figure~\ref{fig:cusp}. This data combination also provides only a very weak constraint on the stellar mass-to-light ratio, as might be expected given the relatively small number of GCs in the centre as compared to the stars. When we compare the predictions of the best-fit model in this case, however, we find that it significantly overpredicts the stellar velocity dispersions in the centre. Adding in the stellar data, then, requires an excavation of the central regions, giving rise to a sub-NFW inner slope more easily reconciled with the assumption of \citet{Zhu2014}, although we do still find a lower stellar mass, which could be a consequence of the fact that, unlike \citet{Zhu2014}, we include the BH in our models. The solution of the discrepancy between the two studies is presumably then that it is simply not possible to constrain the whole mass profile using GC data alone. This really shows the importance of combining multiple tracer populations with different characteristic radii.

We also perform the first joint inference on the BH, stellar and halo mass parameters that has been carried out with the high-resolution NIFS kinematics of the stars in the central 2$''$. When these data were presented in \citet{Gebhardt2011}, the structure and scale of the halo and stellar distributions were fixed at \citet{Murphy2011}'s values, which in turn were inferred assuming the BH mass measured in \citet{Gebhardt2009}, based on lower-resolution data. The BH mass that we infer is consistent with the \citet{Gebhardt2011} value within uncertainties: for the best cgNFW model, we find $7.22_{-0.40}^{+0.34} \times 10^9 M_{\odot}$. 

%One assumption of our anisotropy model that could systematically increase our BH mass is that the stars become isotropic centrally; however, this assumption is motivated by cosmological simulations \citep[e.g.][]{Diaferio1999,  Colin2000, Wojtak2005} and is unlikely to be far from reality. On the other hand, the \citet{Gebhardt2011} model assumed a very high mass-to-light ratio $\Upsilon_{\star,V} = 9.1$, and it may be that covariance between the BH and stellar mass led to a bias in their result. We test this by fixing our halo parameters and stellar mass-to-light ratio at \citet{Murphy2011}'s values -- converted from their assumed distance $D_L$ = 17.9 Mpc to our 16.5 Mpc -- and inferring the BH mass alone, and we do indeed find this to have a significant impact on the BH mass, which drops by 0.15 dex. This lends support to our hypothesis, and highlights the value of jointly modelling the entire mass profile with all its components.

In previous models of M87, significant tension has also arisen between the ratio of the stellar and halo masses and expectations from abundance matching, and this has tended to be related to a consistently high value of the stellar mass-to-light ratio, $\Upsilon_{\star}$. For instance, \citet{Zhu2014} infer a very high mass-to-light ratio $\Upsilon_{\star,V} \sim 7.5$, despite the circular velocity of their LOG halo profile being slightly lower than ours (at $v_c = 591 \pm 50$ kms$^{-1}$ compared to our 610 kms$^{-1}$), and this places it above the abundance matching predictions of \citet{Moster2010}, even allowing for uncertainties. \cite{Murphy2011} infers both a larger stellar mass (with $\Upsilon_{\star,V} = 9.1 \pm 0.2$ at an assumed distance $D_L = 17.9$ Mpc, which translates to $\Upsilon_{\star,V} = 9.9$ at our assumed distance) and a larger halo mass ($\log(M_{vir}/M_{\odot}) = 14.37$, $R_{vir} = 1.6$ Mpc), also placing M87 above abundance matching expectations. While our Nuker profile for the projected light is not strictly normalisable, we compare our results using the absolute magnitude of M87 $M_v = 8.30$ reported in \citet{Kormendy2009}, corrected for dust reddening using the extinction given in that study. This gives a stellar mass $\log(M_{lum}/M_{\odot}) = 11.61 \pm 0.10 $. Compared with the virial mass $\log(M_{vir}/M_{\odot}) = 13.87_{-0.35}^{+0.42}$, this puts M87 just within the abundance-matching curves of \cite{Moster2010}, allowing for uncertainties, and so alleviating this tension. We note, however, that all of our isotropic models -- as well as our anisotropic LOG model -- also yield anomalously high stellar masses relative to the halo, and furthermore that all these previous studies assumed a form for the halo given by the LOG profile, which we find to not perform as well as the more generalised NFW-like parameterisations. Seen together, then, this highlights the need for flexible density and anisotropy models in order to properly understand the astrophysics of these systems.

\subsection{The case for a cored halo}

It is important to compare our findings for the halo structure with other ETGs; as explained in the Introduction, this is crucial for developing our understanding of the role of baryons in shaping the halo and the diversity of structures that can arise. 

A number of other studies have found BCGs to have DM profiles which  are flatter than NFW: \cite{Newman2011}, for instance, combined stellar dynamics with weak and strong graviational lensing and X-ray data to infer a density slope $\gamma < 1$ for the BCG Abell 383 with 95 \% confidence, while \cite{Newman2013} went on to fit gNFW halo models to a sample of seven massive BCGs and found a mean slope of $\gamma = 0.5 \pm 0.1$, inconsistent with the NFW prediction -- and with three of those galaxies hitting the $\gamma = 0$ prior. This is also in line with the earlier studies of \cite{Sand2002, Sand2004, Sand2008, Newman2009}; more recently, \citet{Caminha2015} have also modelled multiple image families lensed by the cluster Abell S1063 to infer a significant core radius $r_c \sim 100$ kpc. The picture, however, is quite different for elliptical galaxies in the field, and a number of strong lensing studies of ETGs have found halo slopes consistent with NFW or even super-NFW DM profiles: \cite{Sonnenfeld2015}, for instance, modelled the global properties of a sample of 81 lenses and found the inner slope to be consistent with an NFW profile, while \cite{Grillo2012} performed a similar analysis on a smaller sample of lenses from the Sloan Lens ACS Survey (SLACS) and found $\gamma = 1.7 \pm 0.5$, which is very inconsistent with what we find here \citep[though see also][for an example of field ellipticals from the SLACS survey that are found to have sub-NFW inner slopes]{Barnabe2013}. At first sight, this might seem to point tentatively to the existence of two very different evolutionary paths for ETGs in these two environments, and the implausibility of a one-size-fits-all DM halo profile: however, the sample size of ETGs for which this analysis has been carried out clearly remains too small for any meanungful conclusions on this issue.

\begin{figure}
  \centering
  \includegraphics[trim=20 0 5 0,clip,width=0.5\textwidth]{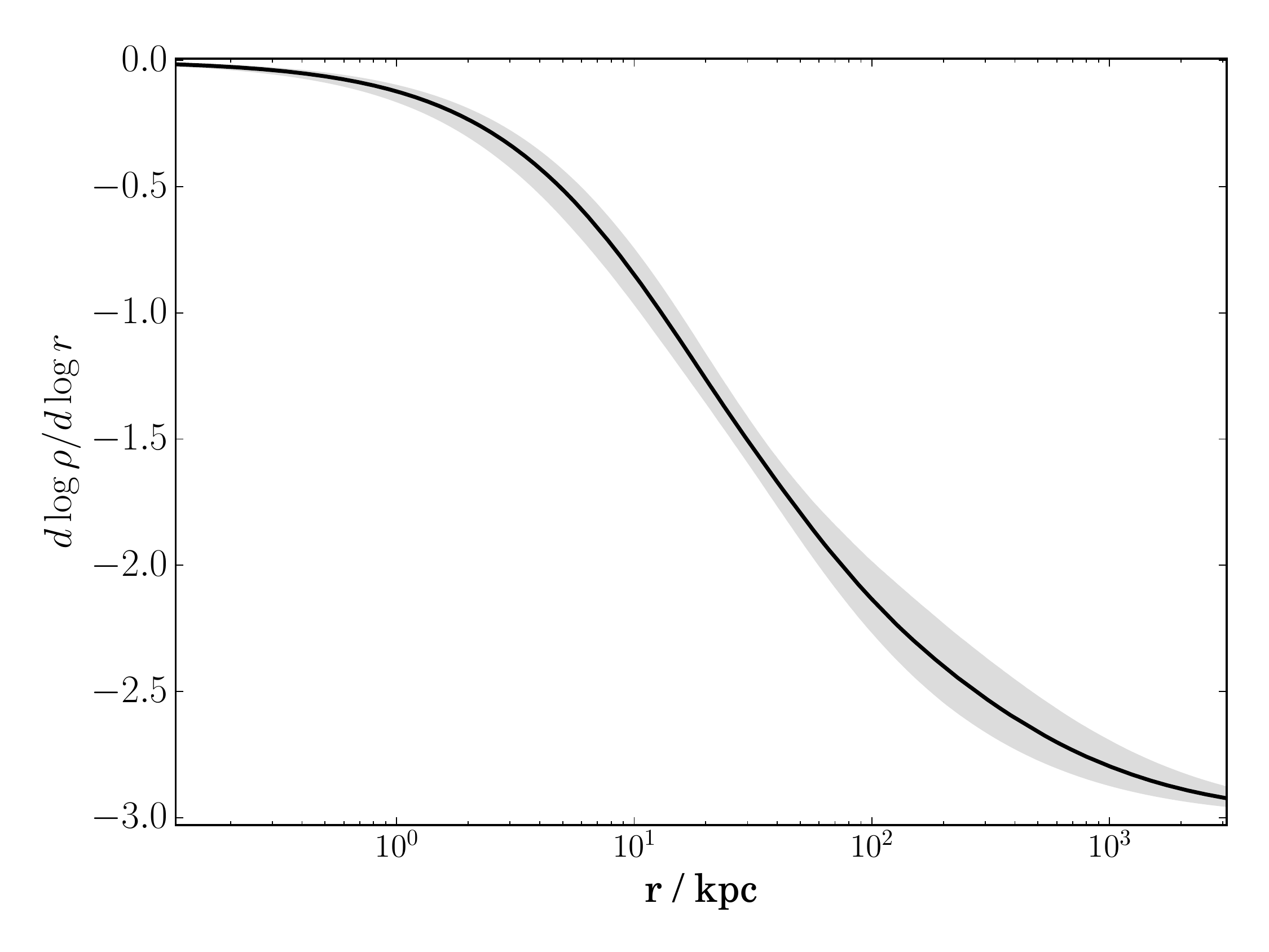}
  \caption{The slope of the halo density profile as a function of radius for the anisotropic cgNFW model. The halo becomes sub-NFW ($\gamma < 1$) at $r \sim 10$ kpc and continues to flatten quite rapidly, with $\gamma \sim 0.6$ by $r \sim 5$ kpc.}
\label{fig:drhodr}
\end{figure}

We now consider the implications of M87's central core in the context of the baryonic feedback processes that play a role in the centres of galaxies. While a cored halo is inconsistent with DM-only simulations, there have been a number of recent simulations in which baryonic effects have been included \citep[e.g.][]{Mead2010, Velliscig2015, Schaller2015}, and we note some similarities with these. Here, we focus on two of the most recent studies of this type in order to elucidate the physics, though we stress that this comparison is by no means exhaustive and a number of similar projects have been carried out. Firstly, we consider the zoom-in simulations of \citet{Laporte2015}, which followed two BCGs from redshift $z = 2$ to $z = 0$ and demonstrated that the effect of repeated dissipationless mergers is to soften the otherwise-NFW-like cusp by $\Delta\gamma \sim 0.3-0.4$ on scales of the stellar half-light radius. In this paradigm, infalling satellite galaxies experience dynamical friction from DM as they move through the halo, and this transfers energy to the DM, causing it to expand and so become less dense in the central regions. While we do not measure a core radius explicitly, in Figure~\ref{fig:drhodr} we plot the density slope as a function of radius, $\text{d}\log\rho/\text{d}\log r$, and see that the slope becomes sub-NFW at radii $r < 10$ kpc and that $\gamma \sim 0.6$ (equivalent to $\Delta\gamma \sim 0.4$) at r$ \sim 5$ kpc, which is comparable to the scale radius of the starlight measured by \cite{Kormendy2009}. However, we find a cored centre while these simulations find only $0.6 < \gamma < 1$ down to the innermost resolvable radius. While they suggest that the merging of SMBHs could then give rise to cores in the central 3-4 kpc -- with the binary BH system spiralling inwards and transferring energy to the surrounding matter -- their simulations do not include the effects of SMBHs and so they are unable to test this. It therefore remains unclear as to whether merging events would be capable of totally erasing the cusp.

On the other hand, M87 has a large AGN at its centre and this may also contribute to core formation. The second study that we examine is that of \cite{Martizzi2012}, which specifically simulated Virgo-like ETGs using a recipe for AGN feedback. They found cores to develop within the inner 10 kpc. In this scheme, outflows of gas due to the AGN are able to irreversibly modify the gravitational potential of the halo and so cause expansion of both the dark and luminous matter. In that study, they suggest that a combination of AGN feedback and SMBH effects are the main contributing processes, though they also note that the large amount of gas expelled by the AGN increases the efficiency of the energy transfer due to dynamical friction. This role for AGN is analogous to the role of supernovae in dwarf spheroids \citep[e.g.][]{Navarro1996, Pontzen2012}. However, while the core radius of 10 kpc seems consistent with the density profile that we infer, \cite{Martizzi2012} predict cores in the 3D stellar density, whereas M87 is known to have a `cuspy core' in its 2D surface brightness ($\gamma = 0.18$), making it considerably cuspy in 3D. Indeed, the finding of cores in the 3D density distribution of the stars seems at odds with observations of large numbers of ETGs \citep[e.g.][]{Kormendy2009}.  Again, this seems to be a sign of the complex nature of the processes governing the shape of the density profile, and that there still remain many questions to be answered on this subject.

\subsection{Implications for the stellar initial mass function}

One of the initial motivations for this investigation was to resolve the striking discrepancy between the $\Upsilon_{\star,V}$ inferred by \citet{Agnello2014} and \citet{Zhu2014}, particularly in light of recent evidence that massive ETGs may have Salpeter-like stellar mass-to-light ratios \citep[e.g.,][]{vanDokkum2010,Auger2010,Cappellari2012}. In our anisotropic cgNFW model, we find $\Upsilon_{\star,V} = 3.50_{-0.36}^{+0.32}$; however, \citet{Zhu2014} infer a much higher value $\Upsilon_{\star,V} \sim 7.5$, and previous studies have found mass-to-light ratios even higher than this \citep[e.g.][who find $\Upsilon_{\star,V} \sim 9.9$ when converted to our assumed distance]{Murphy2011}. We use the single stellar population models from \citet[BC03]{Bruzual2003} to understand the implications of the stellar mass-to-light ratio for the age and IMF of M87.

We measure colours for M87 from the CFHT photometry and compare these with the BC03 models. In particular, we determine the mean surface brightness within a 2$^{\prime\prime}$ annulus at 30$^{\prime\prime}$ (i.e., $\sim$2.5 kpc) from the centre of M87, masking globular clusters and other artefacts, for each of the $ugriz$ filters. We find extinction-corrected $u-g$, $g-r$, $g-i$, and $g-z$ colours of 1.48, 0.67, 1.13, and 1.45, respectively, which agree well with the colours calculated over a much larger aperture from SDSS photometry \citep{brown}. Then, assuming a Salpeter IMF and solar metallicity, the BC03 models give an age of 10 Gyr and $\Upsilon_{*,V}\approx5.7$, significantly lower than the \citet{Zhu2014} or \citet{Murphy2011} values and thus implying the need for an IMF even more extreme than that found by \citet{conroy} if these high mass-to-light ratios are to be explained. If we instead assume a Chabrier IMF and solar metallicity, the BC03 models still find a $\sim10$ Gyr age, but $\Upsilon_{*,V}\approx3.7$, which is in good agreement with our dynamical measurement when we allow for anisotropy. We note, however, that our dynamical inference on $\Upsilon_{V,*}$ is very sensitive to our assumptions about anisotropy. Furthermore, our assumption of solar metallicity -- and, indeed, a constant $\Upsilon_{*,V}$ with radius -- may be inappropriate. For example, \citet{Montes2014} use UV to NIR photometry to demonstrate that M87 has colour gradients, and they interpret these as being the result of a metallicity gradient.

% However,  with the central (inner 1 kpc) region having 1.5 times solar metallicity whilst the data are consistent with solar metallicity at larger radii (including the aperture within which we have measured our photometry).

\subsection{Importance of correctly modelling the underlying tracer populations}

The Jeans equation requires us to know the underlying tracer density for each population, and this, along with the mass profile and anisotropy, determines the velocity dispersion that is measured. Therefore any uncertainty or bias in the parameterisation of the tracer density might manifest itself as uncertainty or bias in the resulting velocity dispersions and, in a study such as this in which the velocity dispersions are the primary tool for the inference, also as uncertainty or bias in the mass profile. A particular worry here is the distributions of the GCs, as the spectroscopic catalogue of \cite{Strader2011} is only a subsample of the wider population and, by virtue of its being selected for spectroscopy, is subject to some non-trivial selection function that may change the apparent distribution from that of the true underlying one. It is therefore important to model the distribution using an independent photometric sample. Indeed, this was the motivation for the initial photometric study of \citet{Oldham2015}. 

To demonstrate the impact on the inference of using the density of the spectroscopic subsample in the Jeans analysis rather than that of the more repsresentative photometric sample, we repeat our analysis using the three-population decomposition given in \cite{Agnello2014}. This decomposition was carried out based solely on the \cite{Strader2011} spectroscopic catalogue, and separates the GCs into a compact red component along with two much more extended blue and intermediate-colour components. Our anaylsis differs from \cite{Agnello2014}'s in that they focussed mainly on virial decompositions and used the GC data exclusively, whereas we use a spherical Jeans analysis and combine the GCs with the stellar kinematics and satellite galaxy datasets. This comparison is not intended as a comment on their work, but merely aims to show that, in this particular analysis, the use of this approximation to the true distribution is important. When we carry out the inference using the three populations, we find that we underpredict the halo mass in the region in which the GCs dominate the fit by $\sim 0.25$ dex. $M_{vir}$ also decreases by $\sim 10 \%$, though the effect is presumably smaller here because the GC data only extend out to a fraction of the virial radius. Unsurprisingly, our inference on the stellar mass-to-light ratio is robust against the GC S\'ersic distributions, as this is mainly determined by the properties of the stellar population, which is fixed by our fit to the surface brightness profile. 

Another possible bias in spherical Jeans modelling could result from any flattening in the potential: indeed, there have been recent suggestions that the shape of a galaxy's DM halo may correlate with the shape of its luminous matter \citep[e.g.][]{Wu2014}, and, while the modelling of \cite{Oldham2015} assumed the GCs to be spherically distributed, they do exhibit some ellipticity at larger radii. However, the effect of this was analysed in \citet{Zhu2014} using a series of axisymmetric Jeans models, and it was found that the impact of any realistic amount of flattening on the mass profile was small, with up to a 10 \% decrease in the mass for GCs with elliptical profiles. This is therefore unlikely to be a major concern here.

%This shows, then, that using the incorrect tracer density for any of the tracer populations has a significant systematic effect on the mass inference.

\subsection{Anisotropy, and a pinch radius from multiple tracer populations}
\label{sec:iso}

As explained in Section~\ref{sec:results}, we incorporated anisotropy in our models by assuming a scaled Osipkov-Merritt profile for the stars and constant anisotropies for the GC populations, and inferred mild radial anisotropies for the three populations, which in turn led to changes in our inference on the mass model relative to the isotropic case. As the inclusion of the anisotropy parameters is considerably more expensive from a computational point of view, it is of interest to investigate ways of obtaining robust mass estimates using models with fixed anisotropy. It is well known \citep[e.g.][]{Wolf2010} that, for the Jeans analysis of a single tracer population, there exists a pinch radius at which the dependence of the inferred mass profile on the anisotropy is minimal. This occurs at approximately the 3D effective radius of the tracer and can be quoted as a robust measure of the enclosed mass -- instead of an entire radial profile -- when it is not possible to infer the anisotropy from the data. However, our case is slightly more complicated than those studied by the previous authors as we are combining three completely independent tracer populations. It is therefore of great interest to test whether the pinch radius is still recovered in this case.

\begin{figure}
 \centering
  \includegraphics[width=0.5\textwidth]{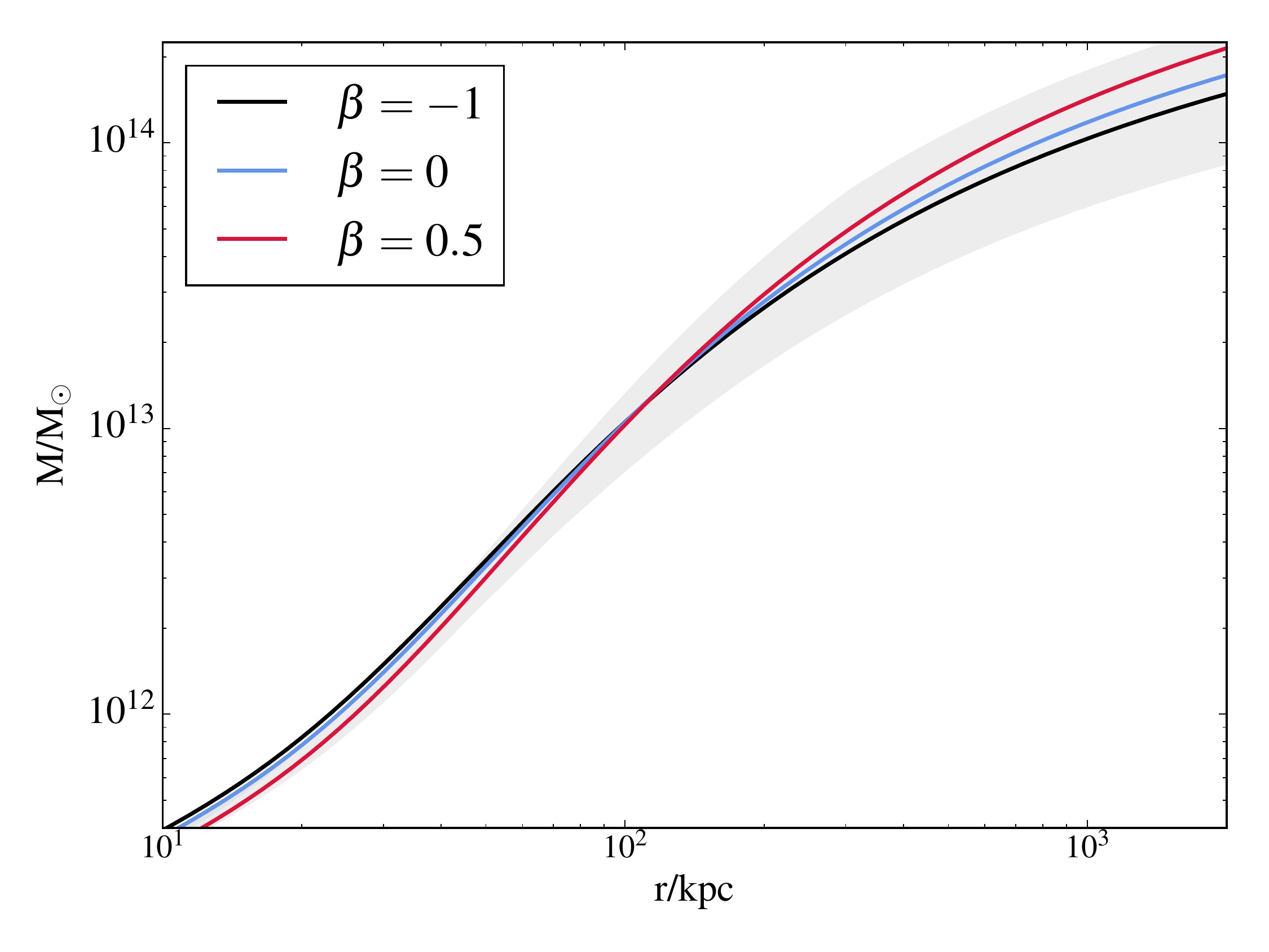} %             #Pinch1.eps}
  \caption{Varying the stellar anisotropy has an effect on the inferred mass profile at all radii except for the pinch radius, clearly visible at $\sim 111$ kpc, where the dependence of the mass on the anisotropy is minimised. While such an extreme tangential bias as $\beta = -1$ is not physically motivated, we run the inference for a range of constant stellar anisotropies between $\beta = -1$ and $\beta = 0.5$ to demonstrate the emergence of the pinch-point.}
\label{fig:pinch}
\end{figure}

For simplicity, we run a series of models in which the stellar anisotropy is fixed at some non-zero constant value while the GCs remain isotropic. This is guided by our earlier results, which found the stars to be considerably more anisotropic than the GCs. To span a range of anisotropies, we let the anisotropy run from $\beta = -1$ (corresponding to the case where the velocity dispersion in the radial direction is twice that in the radial direction) to $\beta = 0.5$ (at which point the tangential velocity dispersion is half the radial velocity dispersion). The results are shown in Figure~\ref{fig:pinch}. Indeed, we do find the three $\beta$-curves to intersect at a single radius, as in the case of single-tracer population models: however, this occurs at a radius $= 111 \pm 4$ kpc (with mass $\log(M(R< R_p)/M_{\odot}) = 13.09 \pm 0.40$), which is much too large to be associated with the effective radius of the starlight. This is presumably a consequence of our use of additional tracer populations, all at larger radii than the stars. This is an interesting finding which will be useful for further multiple-population studies of this kind.

\section{Conclusions}
\label{sec:conclusions}

We have modelled the mass profile of M87 using a combination of stellar, GC and satellite galaxy kinematics in a Jeans analysis, and our main conclusions are as follows:
\begin{enumerate}
 \item M87 is a massive BCG with a sub-NFW DM halo. It has a virial mass of $\log(M_{vir}/M_{\odot}) = 13.87_{-0.35}^{+0.42}$, a virial radius $R_{vir} = 1.1 \pm 0.2$ Mpc and a stellar mass $\log(M_{lum}/M_{\odot}) = 11.61 \pm 0.10$, placing it at the top end of the galaxy mass distribution and just within the stellar-to-halo mass relations expected from abundance matching.
  \item Under assumptions of isotropy, the scatter introduced by the use of different models to parameterise the halo is small. In this paradigm, we find M87 to have a high stellar mass-to-light ratio of $\Upsilon_{\star} = 6.9 \pm 0.1$ in the $V$-band and a cored DM halo, with inner slope $\gamma < 0.14$ at the 95 \% confidence level. 
  \item When the model is modified to allow each tracer population some element of anisotropy, differences arise between the halo models, and a cored generalised NFW profile provides the best description of the dynamics. This reduces the stellar mass-to-light ratio to $\Upsilon_{\star} = 3.50_{-0.36}^{+0.55}$ and slightly relaxes the constraint on the inner slope, with a core radius $r_c = 19.0 \pm 8.3$ kpc. All three tracer populations are characterised by mildly radially anisotropic orbits.
  \item The $V$-band stellar mass-to-light ratio of M87 is consistent with a picture in which its stellar populations are old ($\sim$ 10 Gyrs), with solar metallicity. In the isotropic case, this implies a Salpeter-like IMF, whereas when anisotropy is accounted for, a Chabrier-like IMF is preferred.
  \item The inclusion of tracers at a variety of spatial scales has a significant impact on the inference. Modelling the mass using the GCs and satellite galaxies alone - that is, without the stars - we infer a cuspy halo, with inner slope $\gamma \sim 1.5$, but when the stars are included in the inference, the halo becomes sub-NFW. This shows the importance of consistently modelling the profile across a range of spatial scales.
 \item It is important to properly characterise the distributions of the underlying tracer populations as opposed to those of their kinematic subsamples. We have shown that the use of GC colour and spatial distributions based on the kinematic dataset alone leads to a systematic underprediction of the total mass of the system.
   \item When the inference is carried out with different (constant) values for the stellar anisotropy between -1 and 0.5, a pinch radius clearly emerges at which the dependence of the enclosed mass on the anisotropy is minimised. This gives an estimate of the mass enclosed at 110 kpc as $\log(M(R< 110)/M_{\odot}) = 13.09 \pm 0.40$ which should be robust against assumptions about anisotropy.
\end{enumerate}

\section*{Acknowledgements}
We thank the referee for providing helpful suggestions and corrections. We would also like to thank Wyn Evans, Sergey Koposov, Vasily Belokurov, Jason Sanders and Angus Williams for useful discussions, and Debora Sijacki for helpful comments on the draft of this paper. LJO thanks the Science and Technology Facilities Council (STFC) for the award of a studentship; MWA also acknowledges support from the STFC in the form of an Ernest Rutherford Fellowship.

\clearpage
\onecolumn
\appendix
\section{Anisotropy kernel for the scaled Osipkov-Merritt profile}
For the anisotropy profile of Equation~\ref{eq:scaledOM}, the integrating factor defined by Equation~\ref{eq:IF} is 

\begin{equation}
 f(r) = f(r_i)\Bigg(\frac{r^2 + r_a^2}{r_i^2 + r_a^2}\Bigg)^{\beta_{\infty}}.
\end{equation}
The kernel of Equation~\ref{eq:kernel} is then

\begin{equation}
K_{\beta}(R,r) = \frac{\sqrt{r^2 - R^2}}{r} \Bigg( 1 + \beta_{\infty} \Big( \frac{r^2 + r_a^2}{R^2 + r_a^2}\Big)^{\beta_{\infty}} \Big[2 _2F_1 (\frac{1}{2},\beta_{\infty}; \frac{3}{2}; z) - \frac{3R^2 + 2r_a^2}{R^2 + r_a^2} ~_2F_1(\frac{1}{2},\beta_{\infty}+1; \frac{3}{2};z)\Big]\Big)\Bigg)
\end{equation}
where

\begin{equation}
 z = \frac{R^2 - r^2}{R^2 + r_a^2}.
\end{equation}

\clearpage
\twocolumn

%\bibliographystyle{mn2e}
%\bibliography{BIBX.bbl} % same as BIBX.bbl!

\end{document}